\documentclass[12pt,a4paper]{article}

\usepackage{epsfig}
\usepackage{amsmath,amsfonts,amssymb}
\usepackage{cite}
\usepackage{colortbl}
\usepackage{pifont}

\providecommand{\openone}{\leavevmode\hbox{\small1\kern-3.8pt\normalsize1}}

\newcommand{\ptj}{p_{T\,J}}

\begin{document}

\begin{center}
\begin{Large}
{\bf Running bumps from stealth bosons}
\end{Large}

\vspace{0.5cm}
J.~A.~Aguilar--Saavedra \\
\begin{small}
{ Departamento de F\'{\i}sica Te\'orica y del Cosmos, 
Universidad de Granada, \\ E-18071 Granada, Spain} \\ 

\end{small}
\end{center}

\begin{abstract}
For the `stealth bosons' $S$, light boosted particles with a decay $S \to A A \to q \bar q q \bar q$ into four quarks and reconstructed as a single fat jet, the groomed jet mass has a strong correlation with groomed jet substructure variables. Consequently, the jet mass distribution is strongly affected by the jet substructure selection cuts when applied on the groomed jet. We illustrate this fact by recasting a CMS search for low-mass dijet resonances and show a few representative examples. The mass distributions exhibit narrow and wide bumps at several locations in the 100 -- 300 GeV range, between the masses of the daughter particles $A$ and the parent particle $S$, depending on the jet substructure selection. This striking observation introduces several caveats when interpreting and comparing experimental results, for the case of non-standard signatures. The possibility that a single boosted particle decaying hadronically produces multiple  bumps, at quite different jet masses, and depending on the event selection, brings the anomaly chasing game to the next level. 
\end{abstract}


\section{Introduction}

New particles are commonly searched for as a bump in a distribution, sticking out over a smooth background from standard model (SM) processes. The location of the bump either corresponds to the new particle mass, or has a close relation with it. Searches by the ATLAS and CMS experiments at the Large Hadron Collider (LHC) routinely find bumps, of moderate statistical significance, at various locations in the relevant mass distributions. (None of these bumps has been confirmed as a new particle, unfortunately, except the Higgs boson~\cite{Aad:2012tfa,Chatrchyan:2012xdj}.) By construction, should any of these analyses find the particle they seek, its `mass', namely the location of the bump, would roughly be the same independently of the particular event selection applied ---of course, with the height and significance of the bump depending on the signal sensitivity optimisation. This happens because the analyses are designed and calibrated for the specific signals investigated. Still, it is very interesting and pertinent to ask the question whether it would be possible that some other signal might produce bumps at quite different locations other than the true mass, maybe depending on the event selection.

For simple signals, especially those involving charged leptons or photons, that possibility is highly unlikely. But complex hadronic signals can be quite tricky. In previous work~\cite{Aguilar-Saavedra:2017zuc} we have introduced the `stealth bosons', relatively light boosted particles with a cascade decay $S \to A A \to q \bar q q \bar q$, mediated by intermediate particles $A$ (which may not be the same) and decaying into four quarks, which are reconstructed as a single fat jet. Compared for example to boosted weak bosons $W$, $Z$, which give two-pronged jets, the four-pronged jets from stealth bosons have two conspicuous properties: 
\begin{enumerate}
\item For jet substructure variables such as $\tau_{21}$~\cite{Thaler:2010tr,Thaler:2011gf} and $D_2$~\cite{Larkoski:2014gra}, designed to separate hadronically-decaying weak bosons from the QCD background, stealth bosons look more like the QCD background, composed by quark and gluon jets. As we see in the following, the same holds for other proposals~\cite{Moult:2016cvt}.
\item Standard grooming algorithms~\cite{Krohn:2009th,Ellis:2009me,Larkoski:2014wba}, with the usual parameter choices optimised for weak bosons, spoil the jet mass distributions to varying degrees and do not recover the mass of the originating particle, in this case the stealth boson mass. (Of course, a less aggresive grooming attenuates this effect.)
\end{enumerate}
Both facts have already been pointed out previously~\cite{Aguilar-Saavedra:2017zuc}. The goal of the present paper is to study their interplay, which is quite subtle, yet it can easily be understood. Let us consider the decay $S \to A A \to qqqq$ of a boosted stealth boson. When the groomed jet mass $m_J$  is close to $M_S$, the jet substructure is mostly four-pronged, so that a tight requirement of a small $\tau_{21}$ or $D_2$ of the groomed jet, to select a two-pronged substructure, usually results in a rejection. But often the grooming algorithm fully eliminates one of the daughter particles $A$ from the jet, yielding a jet mass $m_J \sim M_A$. This groomed jet has a mostly two-pronged substructure, so the application of a requirement on $\tau_{21}$ or $D_2$ has a much larger efficiency. As a consequence, the bulk of the jet mass bump moves from $M_S$ to $M_A$ after the application of the jet substructure requirement, with the removal of the events with jet mass closer to $M_S$.

We begin by describing in section~\ref{sec:2} our analysis framework, which is a recast of a search for low-mass dijet resonances by the CMS Collaboration~\cite{Sirunyan:2017nvi} that uses a mass-decorrelated jet tagger using the $N_2^1$ variable~\cite{Moult:2016cvt}. The mass decorrelation means that, by construction, the tagging efficiency for the QCD background does not depend on the jet mass, so that the application of a cut on $N_2^1$ does not shape the background. 
Therefore, this experimental analysis is ideally suited for our purpose. In section~\ref{sec:3} we simulate some stealth boson signals and show the `bump running' effect, as the selection on $N_2^1$ is changed. This is already a striking effect, as it may lead to mistaking the identity of a new particle, but also has some other direct consequences that are examined in section~\ref{sec:4}. We discuss our results in section~\ref{sec:5}. Appendix~\ref{sec:a} is devoted to the comparison between jet substructure variables for groomed and ungroomed jets. In appendix~\ref{sec:b} we investigate the effect on the jet mass distributions of a milder grooming, by varying the parameters in the algorithm.

\section{Analysis framework}
\label{sec:2}

\subsection{Signal and background simulation}

The various processes used in this analysis are generated using {\scshape MadGraph5}~\cite{Alwall:2014hca}, followed by hadronisation and parton showering with {\scshape Pythia~8}~\cite{Sjostrand:2007gs} and detector simulation using {\scshape Delphes 3.4}~\cite{deFavereau:2013fsa}. For the signal processes the relevant Lagrangian is implemented in {\scshape Feynrules}~\cite{Alloul:2013bka} and interfaced to {\scshape MadGraph5} using the universal Feynrules output~\cite{Degrande:2011ua}. We use three representative examples,
\begin{align}
& p p \to Z' \to H_1^0 \, Z (\to \nu \nu) \,, && H_1^0  \to A^0 A^0 \,, \notag \\ 
& p p \to Z' \to H_1^0 \, Z (\to \nu \nu) \,, && H_1^0 \to W^+ W^- \,, \notag \\
& p p \to Z' \to H_1^0 \, Z (\to \nu \nu) \,, && H_1^0 \to A_1^0 A_2^0 \,,
\label{ec:signals}
\end{align}
with $S \equiv H_1^0$ a heavy scalar and $A^0$, $A_1^0$, $A_2^0$ pseudo-scalars. In all cases we set the $Z'$ mass to 2.2 TeV. As background processes we consider QCD dijet production and $Wj$, $Zj$ production, with $j$ a light jet. In order to populate with sufficient Monte Carlo statistics the entire mass and transverse momentum range under consideration, we split the samples in 100 GeV slices in the transverse momentum of the leading jet, from 300 GeV to 1 TeV and above, generating $8\times 10^5$ events for QCD dijets, $5 \times 10^4$ events for $Wj$ and $5 \times 10^4$ events for $Zj$ in each slice. The different samples are then recombined with weights proportional to the cross sections. Even if $Wj$ and $Zj$ are sub-dominant, they are included as they produce small bumps in the jet mass distribution at $m_J \sim M_{W,Z}$.

\subsection{Decorrelated jet tagger}

To follow the analysis in Ref.~\cite{Sirunyan:2017nvi}, we select fat jets reconstructed with the anti-$k_T$ algorithm~\cite{Cacciari:2008gp} with radius $R=0.8$, referred to as AK8 jets. Events are selected if they have at least one AK8 jet with transverse momentum $\ptj > 500$ GeV and pseudo-rapidity $|\eta| < 2.5$. The leading jet is the one considered for the analysis. Jets are groomed using the soft-drop algorithm~\cite{Larkoski:2014wba}, with the parameters $z_\text{cut} = 0.1$, $\beta = 0$, which correspond to the modified mass-drop tagger~\cite{Dasgupta:2013ihk}. The $N_2^1$ variable~\cite{Moult:2016cvt} is used to discriminate the two-pronged jets from boosted $Z'$ decays from the QCD background. The jet reconstruction, grooming and jet substructure analyses are performed using {\scshape FastJet}~\cite{Cacciari:2011ma}.

\begin{figure}[t]
\begin{center}
\begin{tabular}{ccc}
\includegraphics[height=3.8cm,clip=]{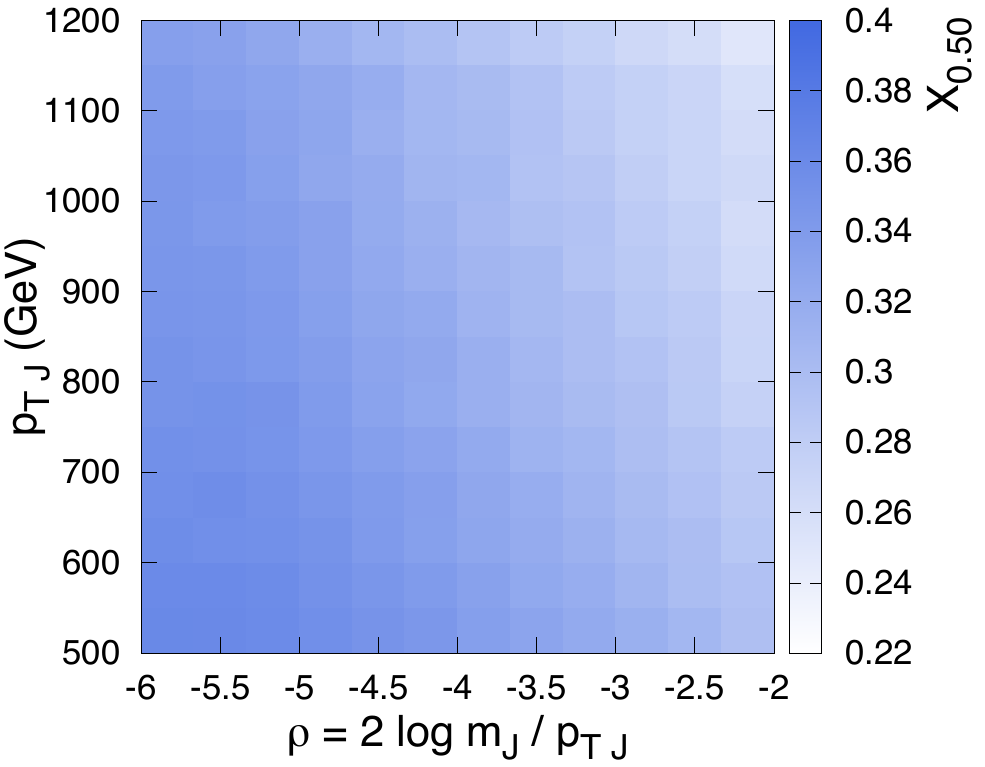} & 
\includegraphics[height=3.8cm,clip=]{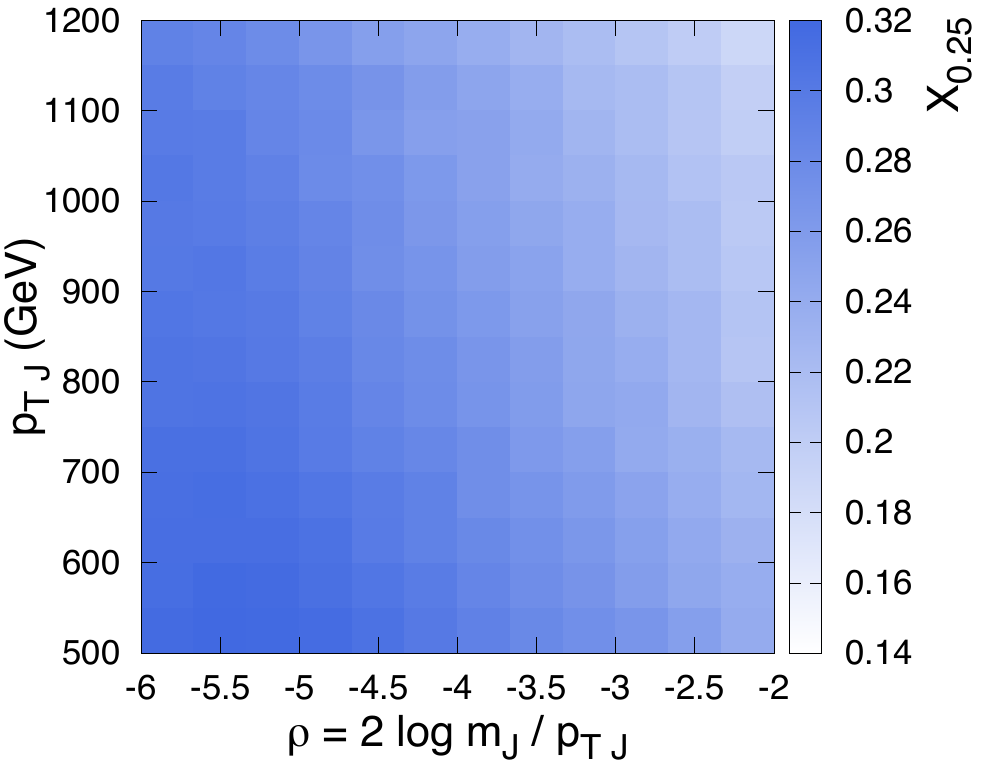} &
\includegraphics[height=3.8cm,clip=]{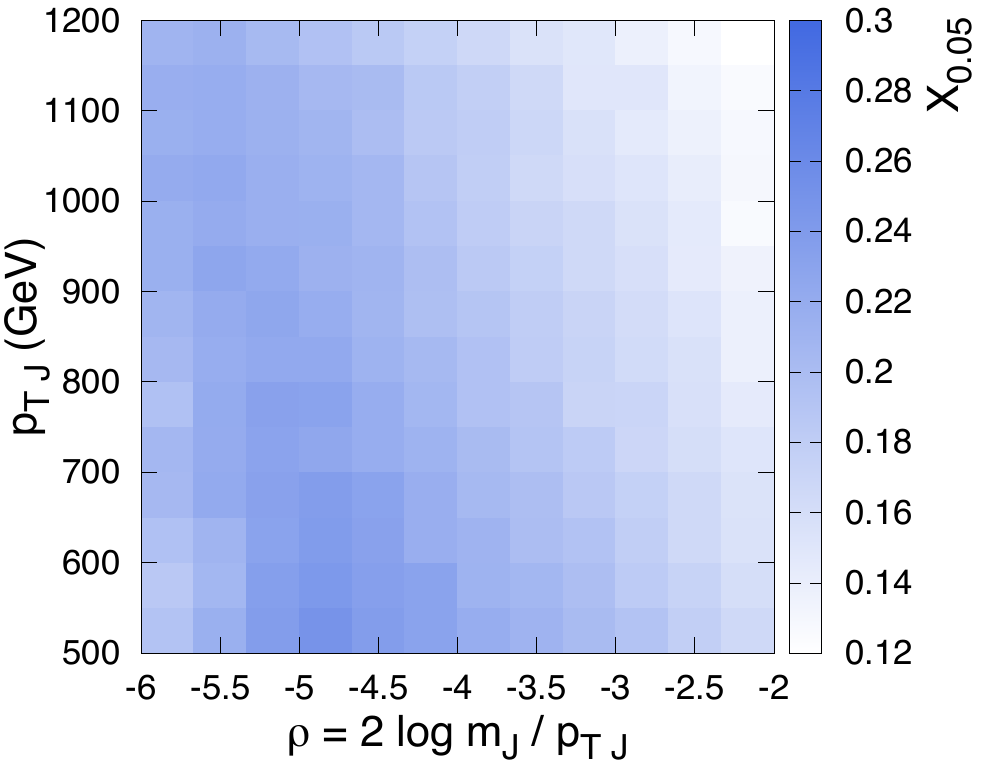} 
\end{tabular}
\caption{Dependence on the jet mass and $\rho$ of the thresholds $X$ corresponding to efficiencies of 50\% (left), 25\% (middle) and 5\% (right) for the QCD background.}
\label{fig:X}
\end{center}
\end{figure}
\begin{figure}[t]
\begin{center}
\includegraphics[height=5.5cm,clip=]{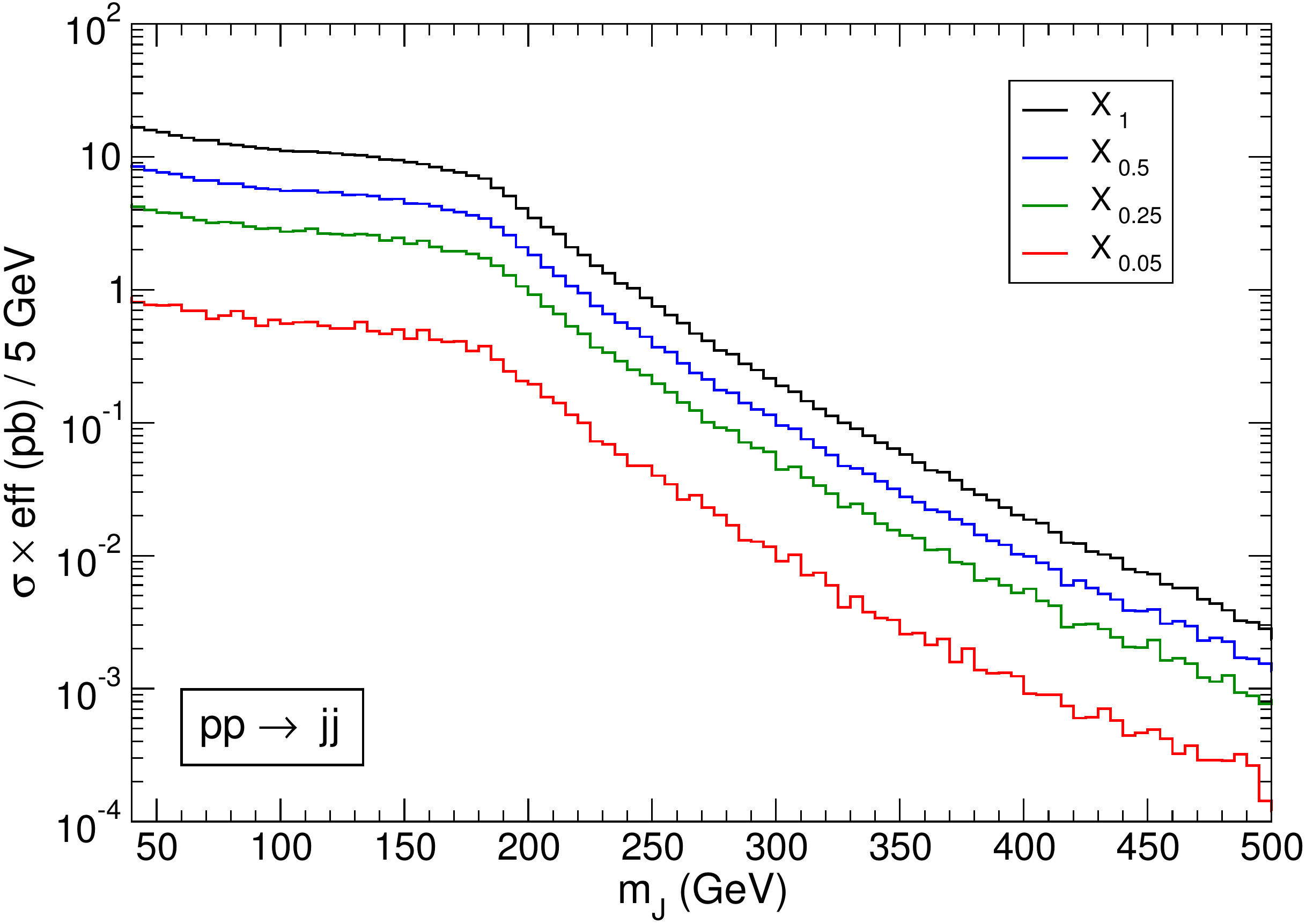} 
\caption{Jet mass spectrum for the QCD dijet background after event selection, without the $N_2^1$ requirement (labelled as $X_{1}$) and with selections corresponding to efficiencies of 50\%, 25\% and 5\%. }
\label{fig:Mincl}
\end{center}
\end{figure}

In order to keep the shape of the jet mass spectrum after the application of a cut on $N_2^1$, a decorrelation method is applied, by varying the cut threshold depending on $\ptj$ and the scaling variable $\rho = 2 \log (m_J / \ptj)$, with $m_J$ the groomed jet mass, keeping a constant efficiency for the QCD background,
\begin{equation}
N_2^1(m_J,\rho) < X(m_J,\rho) \,,
\end{equation}
with $X$ the varying threshold. We consider jets with $-6 < \rho < -2$ and select three sets, $X_{0.50}$, $X_{0.25}$ and $X_{0.05}$, corresponding to working points of 50\%, 25\% and 5\% efficiencies for the QCD background. (The latter is the one used by the CMS Collaboration in their event selection.) The variation of the thresholds with jet mass and $\rho$ is shown in Fig.~\ref{fig:X}.
By comparing with the results in Ref.~\cite{Sirunyan:2017nvi}, one can see that for a 5\% efficiency the thresholds are similar to the ones obtained by the CMS Collaboration. We show the jet mass distribution for QCD dijet production in Fig.~\ref{fig:Mincl}, before the $N_2^1$ cut and with the three selected efficiencies of 50\%, 25\% and 5\%.
We observe that indeed the background is not shaped by the decorrelated $N_2^1$ selection. A kink appears in the distributions at $R \sim 2 m_J / p_{T\,J}$ when the AK8 jet is on the edge of not containing all the jet decay products. The overall normalisation of the background agrees well with CMS measured data~\cite{Sirunyan:2017nvi}, therefore we do not introduce any scaling factor in our simulation.

\section{Running bumps}
\label{sec:3}

We illustrate the running of the bumps when a cut on $N_2^1$ is applied by selecting three stealth boson scenarios.
The first scenario we consider is a stealth boson decaying $S \to AA \to b \bar b b \bar b$, as studied in Ref.~\cite{Aguilar-Saavedra:2017zuc}. Here we choose higher masses $M_S = 300$ GeV, $M_A = 80$ GeV in order to show the effect more clearly. (For the mass values considered in Ref.~\cite{Aguilar-Saavedra:2017zuc}, the displacement of the bumps is around 20 GeV.) This type of signal can take place in left-right models, with $S=H_1^0$ the heavy scalar produced from the decay of a heavier $Z'$ or $W'$ boson and $A=A^0$ the pseudo-scalar in the bidoublet~\cite{Aguilar-Saavedra:2015iew}. The second scenario is $S \to WW \to q \bar q q \bar q$, with $M_S = 300$ GeV, and is used to test possible differences between light quarks $q$ and $b$ quarks. The decay $S \to ZZ \to q \bar q q \bar q$ is analogous. Those signals can also appear in left-right models if the neutral scalar sector departs from the alignment limit, and in models with warped extra dimensions, with $S = \phi$ the radion~\cite{Agashe:2016kfr,Agashe:2017wss}. The third scenario is $S \to A_1 A_2 \to b \bar b b \bar b$, with $M_S = 200$ GeV and two different \mbox{(pseudo-)}scalars $A_1$, $A_2$, with $M_{A_1} = 20$ GeV, $M_{A_2} = 115$ GeV. This type of decay is possible in models with an extended scalar sector, in particular in supersymmetry~\cite{Fidalgo:2011ky,Ellwanger:2017skc}, and we use it to illustrate what happens when there is a hierarchy between the masses of the two decay products of $S$. Another possibility studied in Ref.~\cite{Aguilar-Saavedra:2017zuc} is $S \to Z A$, which can also appear in left-right models. A detailed discussion is omitted here for brevity, as it produces results similar to the cases studied.

\begin{figure}[htb]
\begin{center}
\includegraphics[height=5.5cm,clip=]{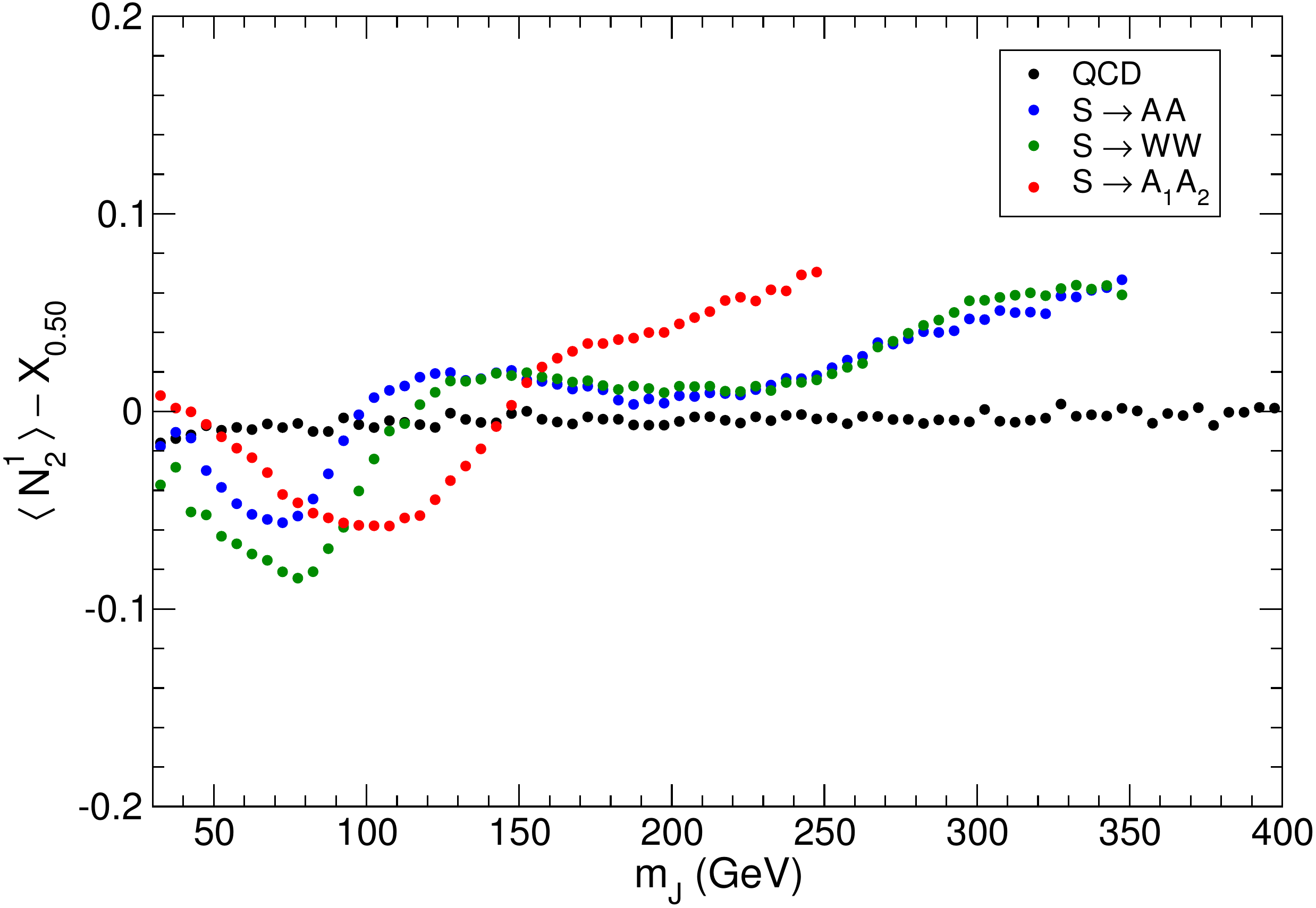} 
\caption{Dependence on the jet mass of the decorrelated average $\langle N_2^1 \rangle -X_{0.50}$.}
\label{fig:avgN2}
\end{center}
\end{figure}

The effect of the jet grooming on the two-subjettiness, as measured by $N_2^1$, can be understood by considering the decorrelated average $\langle N_2^1 \rangle - X_{0.50}$. This quantity is presented in Fig.~\ref{fig:avgN2} for the QCD background and the three scenarios considered. 
(For QCD, the decorrelated average is slightly different from zero because we compute the average, not the median.) It is clearly seen that a requirement of small $N_2^1$ favours lower jet masses: notice that the dips of these distributions are precisely at the masses of the daughter resonances, $M_A$, $M_W$ or $M_{A_2}$, strongly suppressing events with a jet mass near $M_S$. These pronounced dips do not appear when $N_2^1$ for the ungroomed jet is considered in the analysis. A comparison of $N_2^1$ for groomed and ungroomed jets, and their dependence on the transverse momentum, is given in appendix~\ref{sec:a}.

We examine how the bump running effect would show up by adding the three above signals to the SM background. We apply the event selection criteria of the CMS analysis~\cite{Sirunyan:2017nvi} and consider the leading jet mass distribution. Because the signals have large transverse momentum, we also require $\ptj > 900$ GeV for the leading jet, for both the signal and the background. The cross section times efficiency of the injected signals is given in Table~\ref{tab:xsec}. We note in passing that, as previously seen for the $\tau_{21}$ and $D_2$ variables~\cite{Aguilar-Saavedra:2017zuc,Aguilar-Saavedra:2017rzt}, the efficiency for stealth bosons of a cut on two-subjettiness, as measured by $N_2^1$, is smaller than for the QCD background.

\begin{table}[htb]
\begin{center}
\begin{tabular}{lcccc}
& $\sigma \times \text{eff}\;(X_1)$ & $\text{eff}\; X_{0.50}$   & $\text{eff}\; X_{0.25}$  & $\text{eff}\; X_{0.05}$  \\
$S \to AA$         & 94 fb &   0.29  & 0.086 & 0.015 \\
$S \to WW$       & 105 fb & 0.28  & 0.095 & 0.029 \\
$S \to A_1 A_2$ & 64 fb & 0.36 & 0.18 & 0.025
\end{tabular}
\caption{Cross section times efficiency (without the $N_2^1$ selection) for the injected signals, and efficiency of the various $N_2^1$ selection thresholds. \label{tab:xsec} }
\end{center}
\end{table}

\begin{figure}[t]
\begin{center}
\begin{tabular}{cc}
\includegraphics[height=5.5cm,clip=]{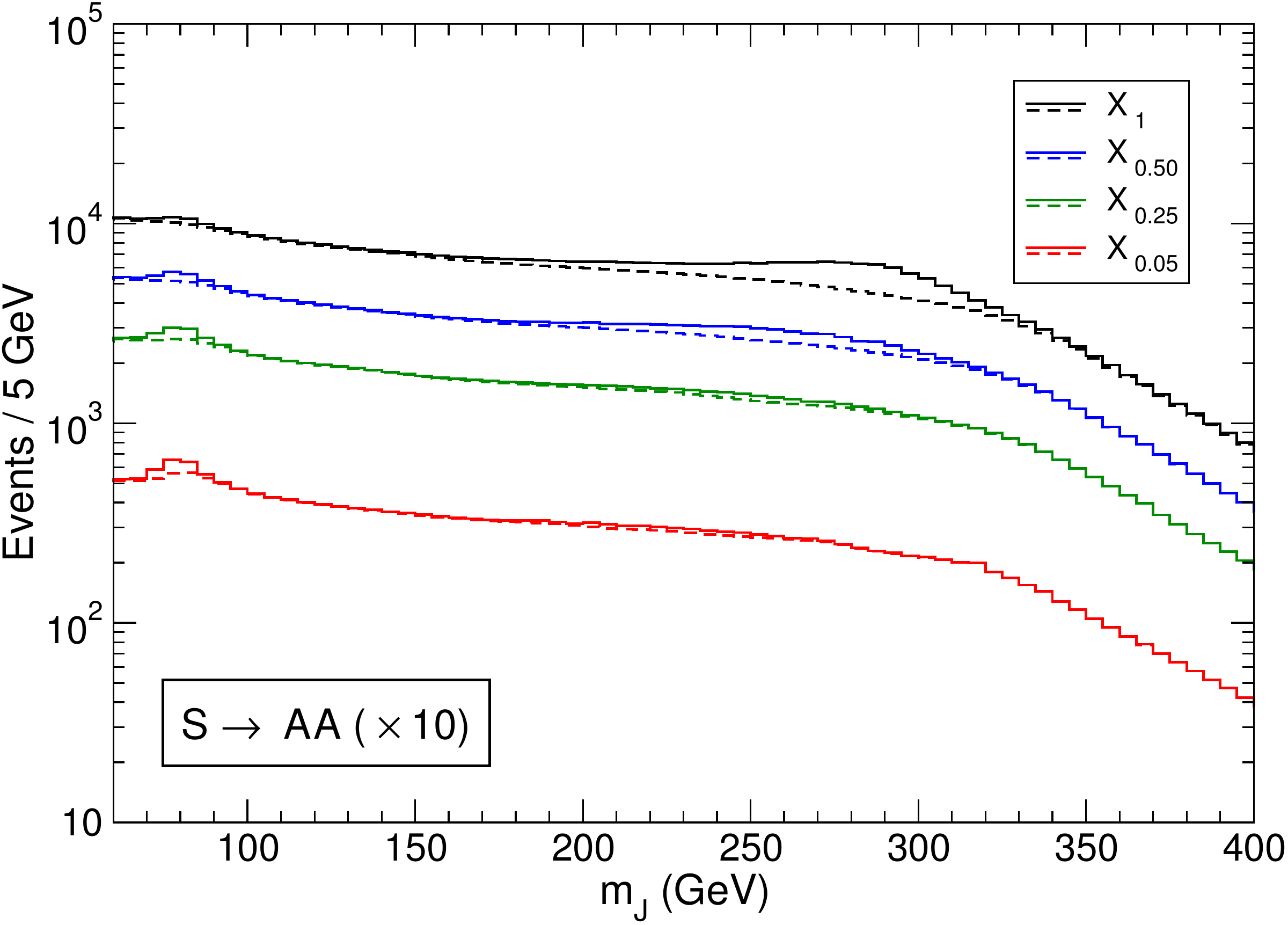} & 
\includegraphics[height=5.5cm,clip=]{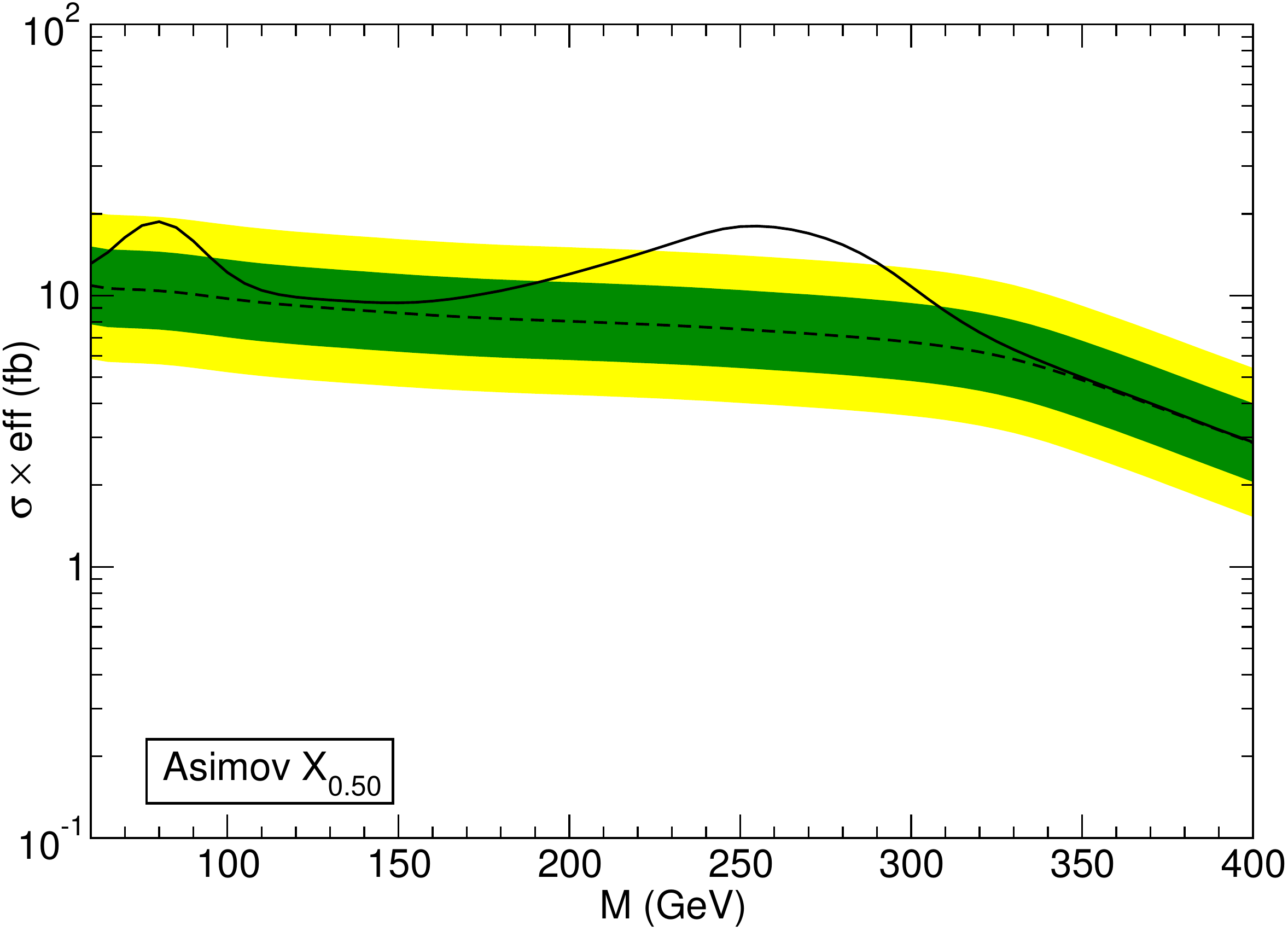} \\
\includegraphics[height=5.5cm,clip=]{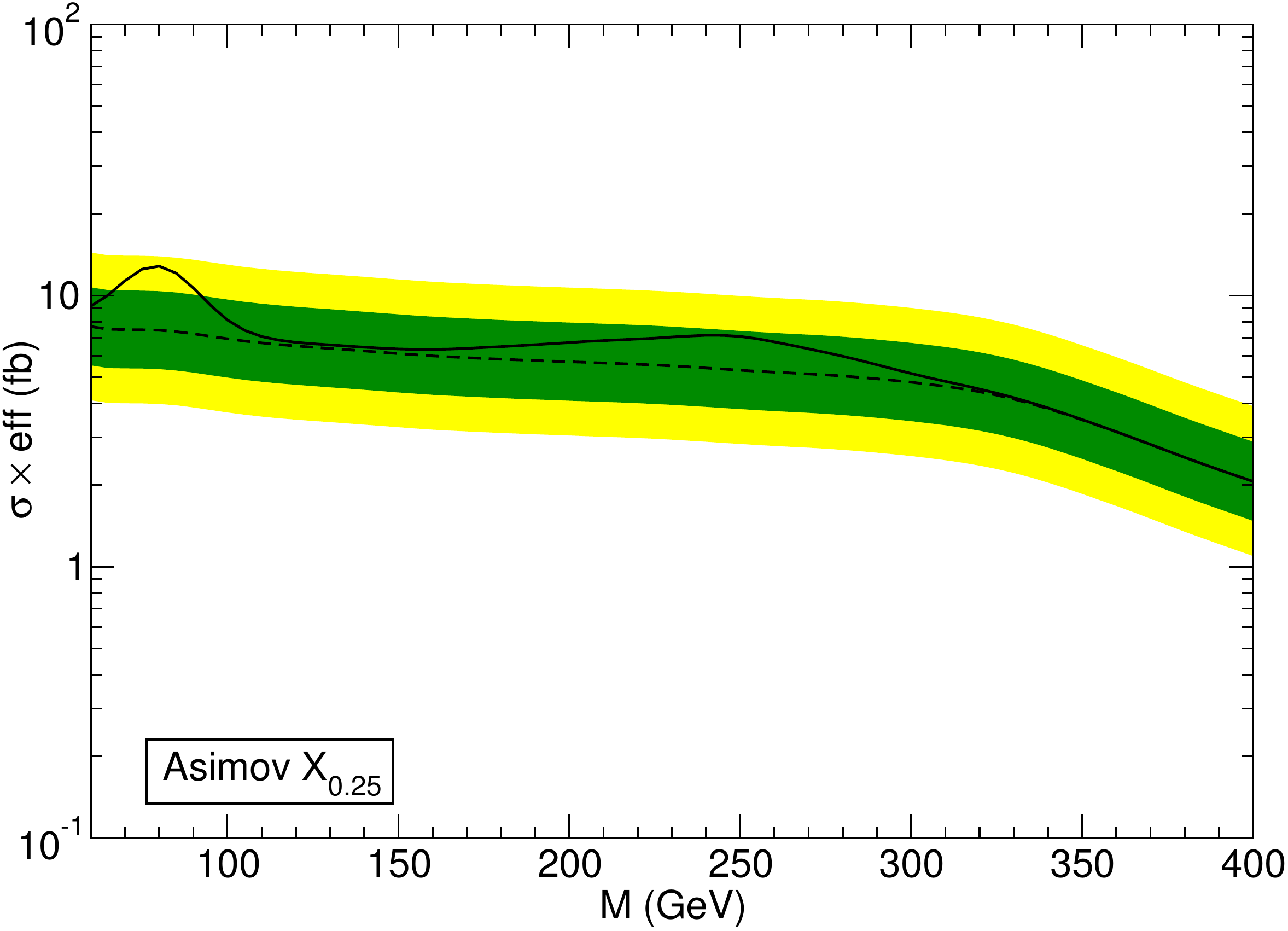}  &
\includegraphics[height=5.5cm,clip=]{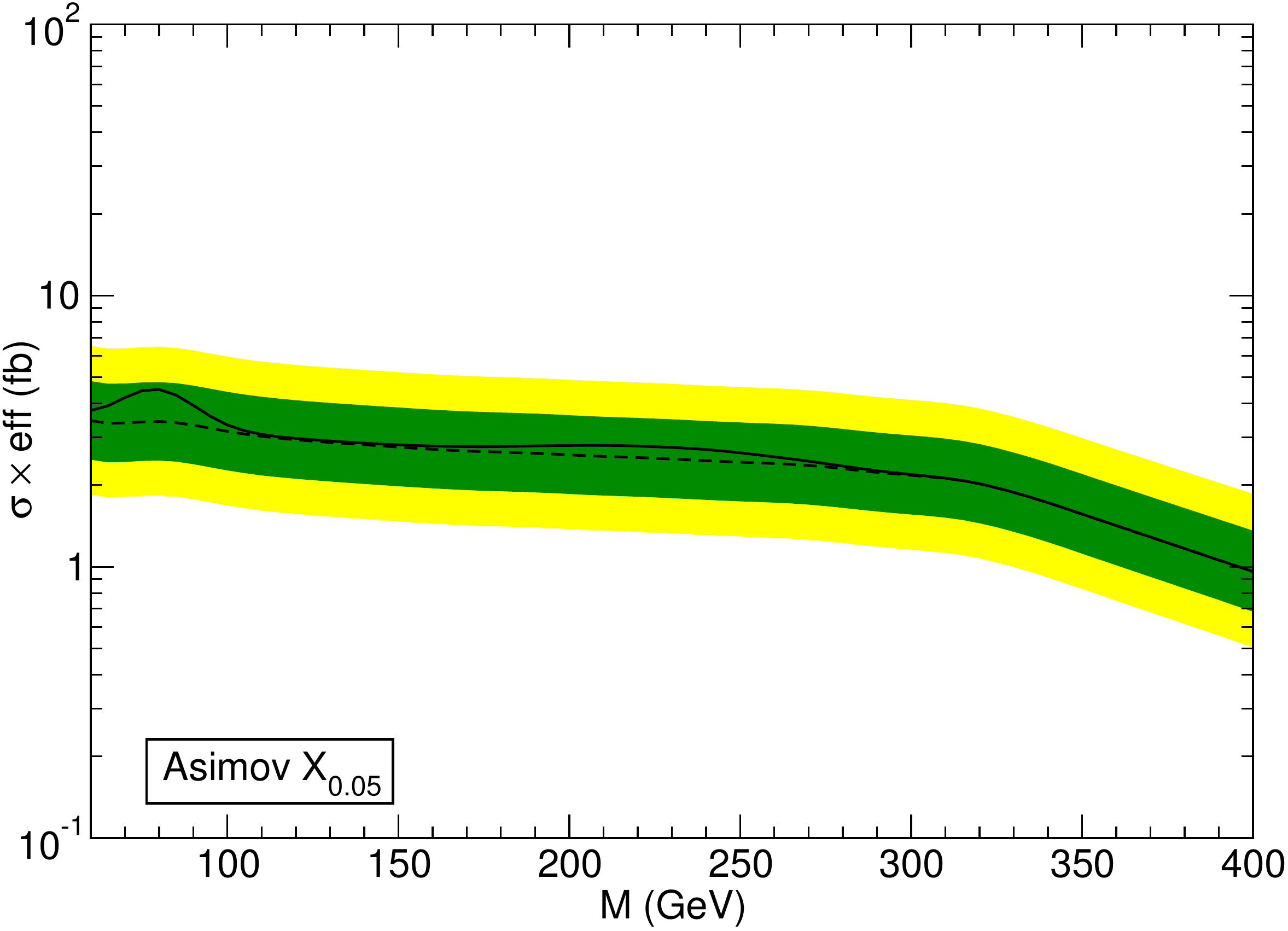}
\end{tabular}
\caption{Jet mass distributions and observed limits for the $S \to AA$ scenario. Top, left: mass distributions with an injected signal, without a cut on $N_2^1$ and with cuts corresponding to the three working points (see the text). Top, right and bottom: expected and observed limits on narrow resonances corresponding to the signal injected, in the three working points for the $N_2^1$ selection. }
\label{fig:AA80}
\end{center}
\end{figure}

\begin{figure}[t]
\begin{center}
\begin{tabular}{cc}
\includegraphics[height=5.5cm,clip=]{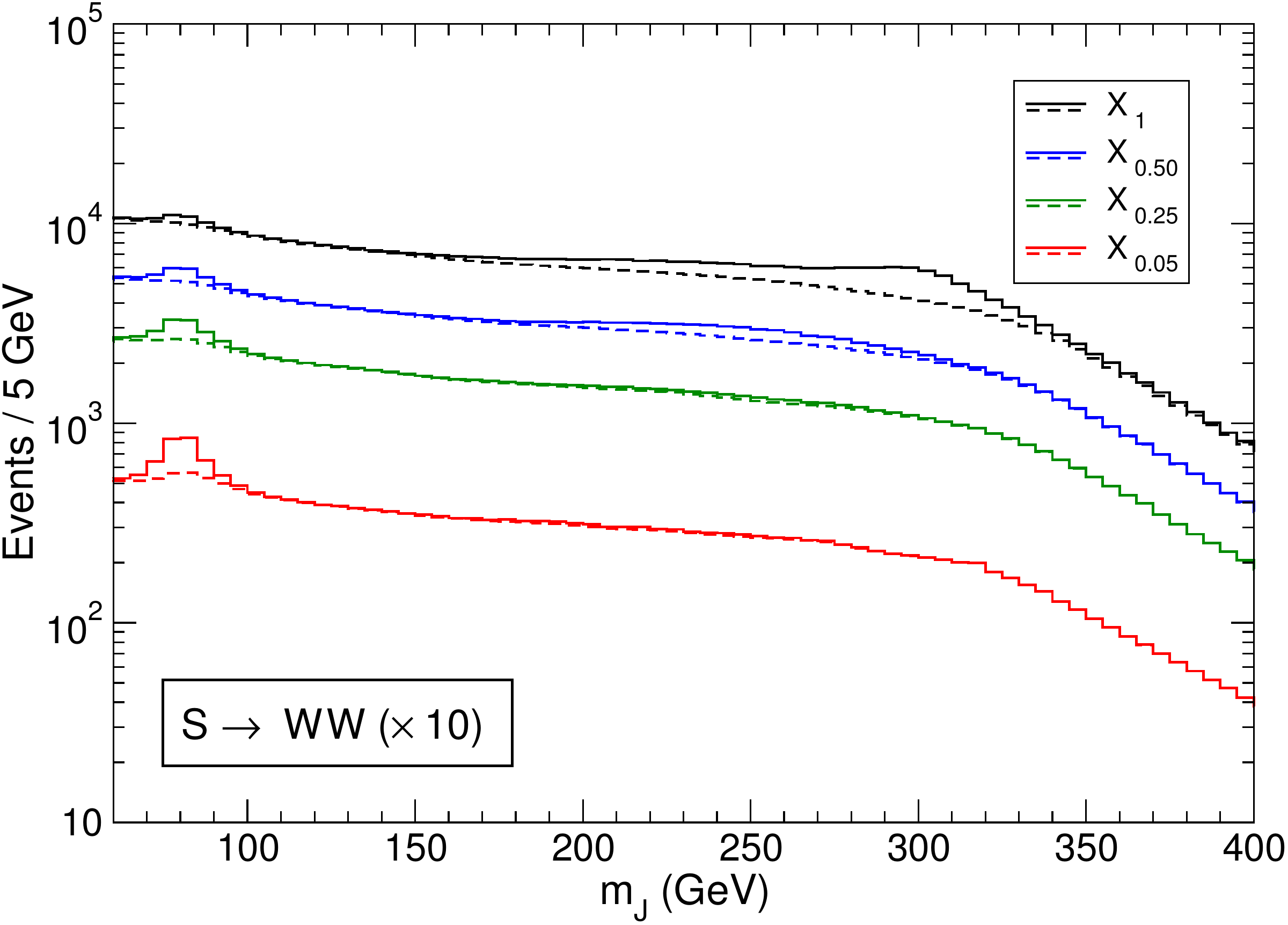} & 
\includegraphics[height=5.5cm,clip=]{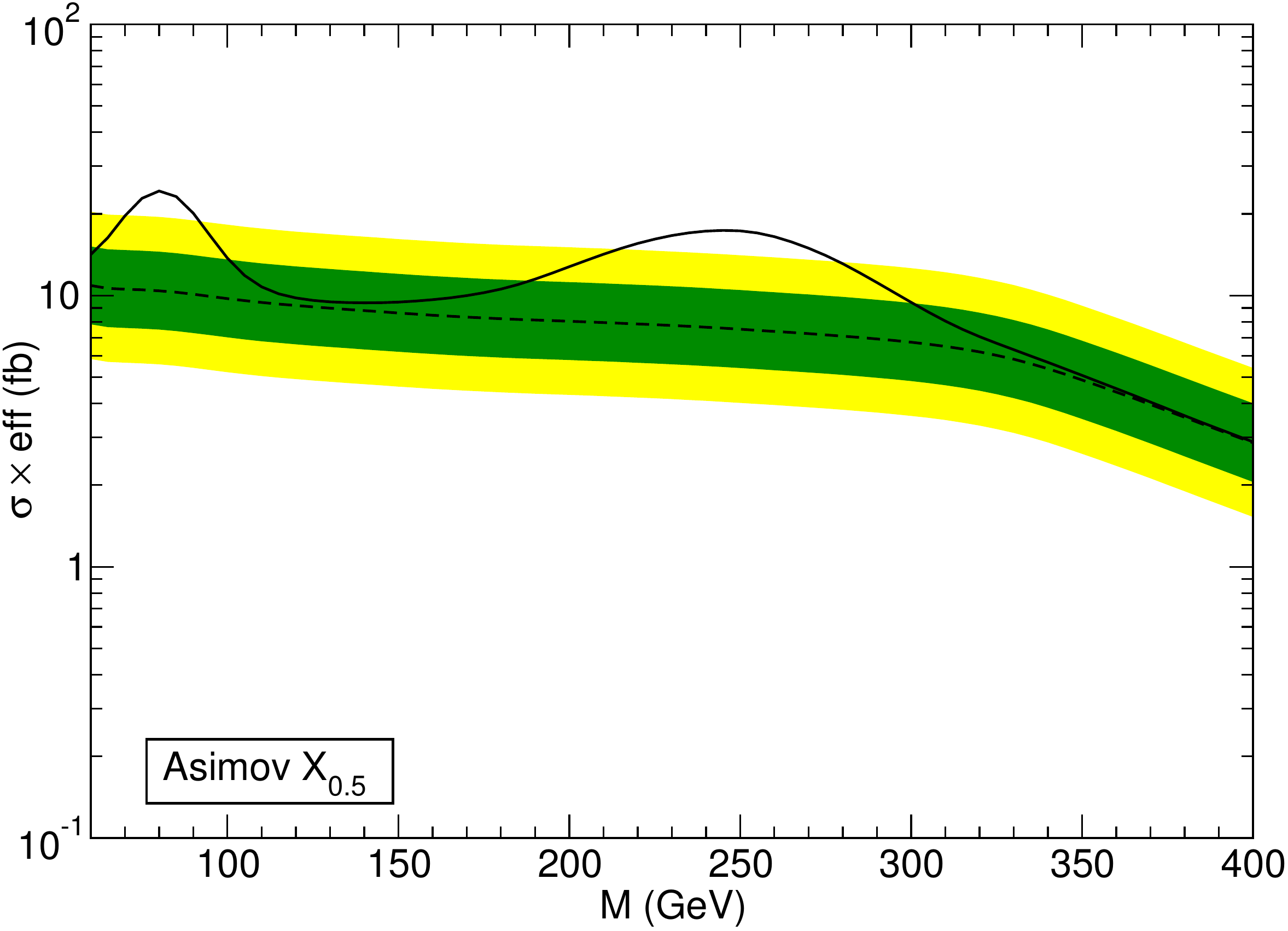} \\
\includegraphics[height=5.5cm,clip=]{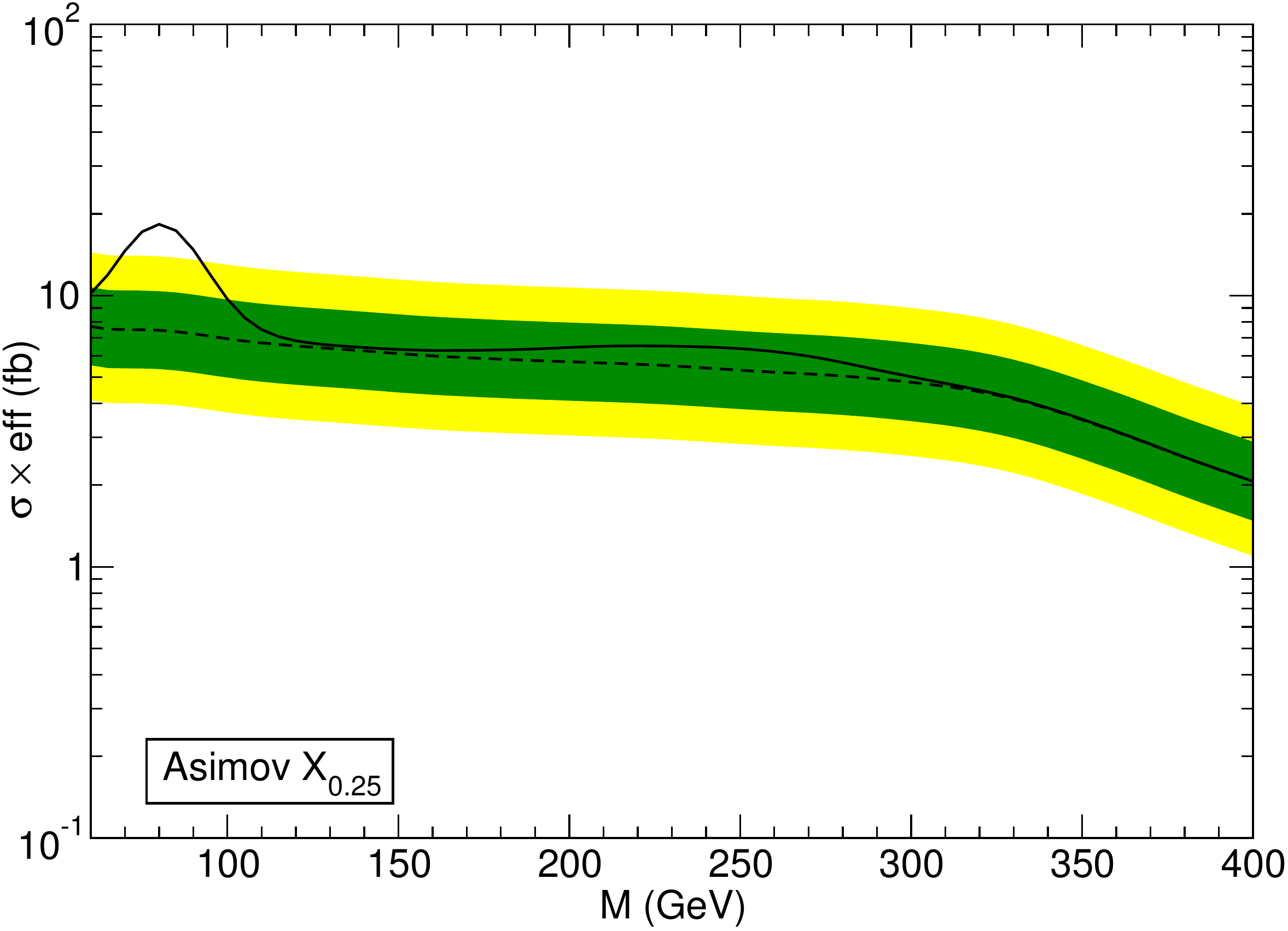}  &
\includegraphics[height=5.5cm,clip=]{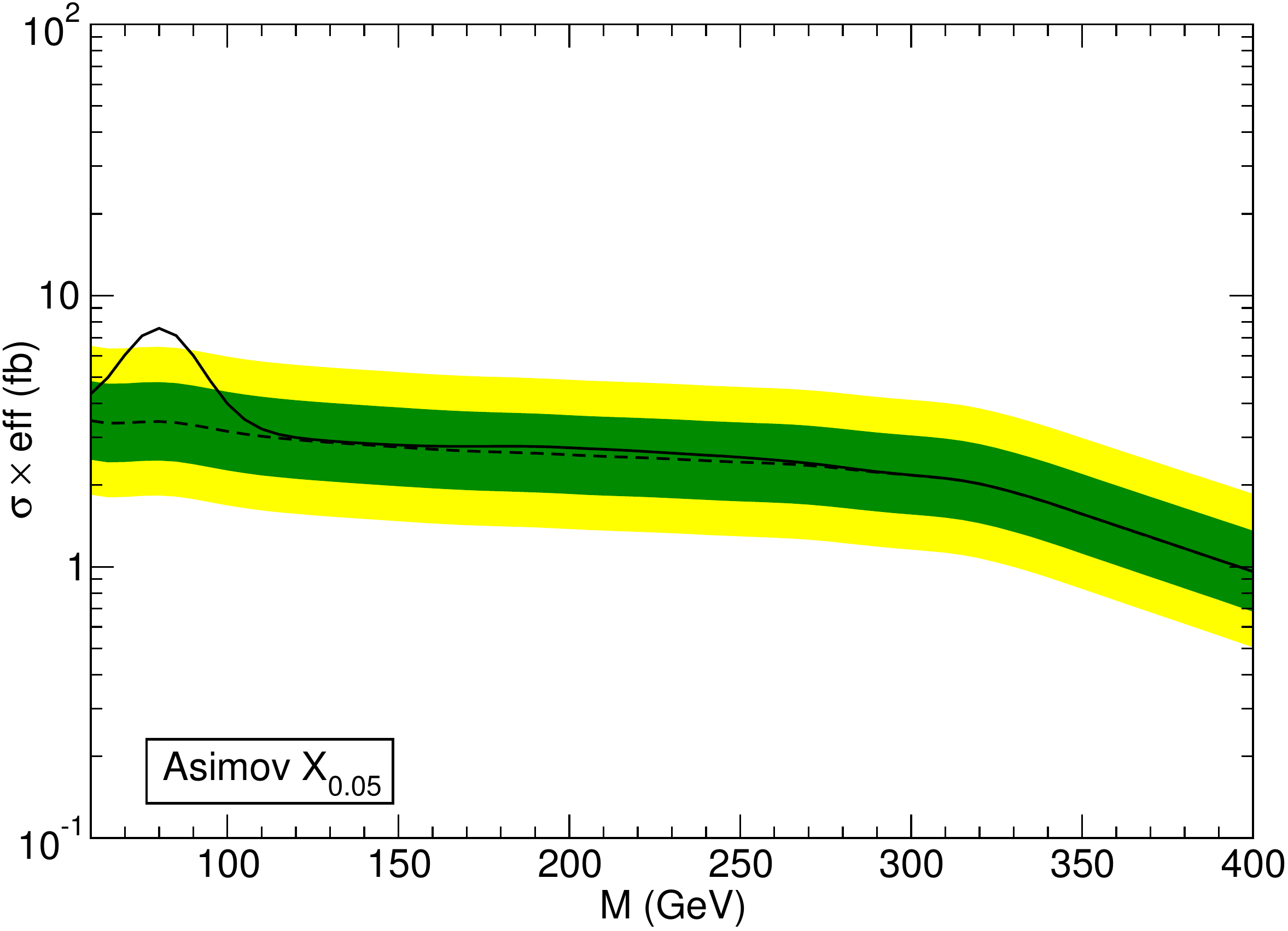}
\end{tabular}
\caption{The same as Fig.~\ref{fig:AA80}, for the $S \to WW$ scenario.}
\label{fig:WW}
\end{center}
\end{figure}

\begin{figure}[t]
\begin{center}
\begin{tabular}{cc}
\includegraphics[height=5.5cm,clip=]{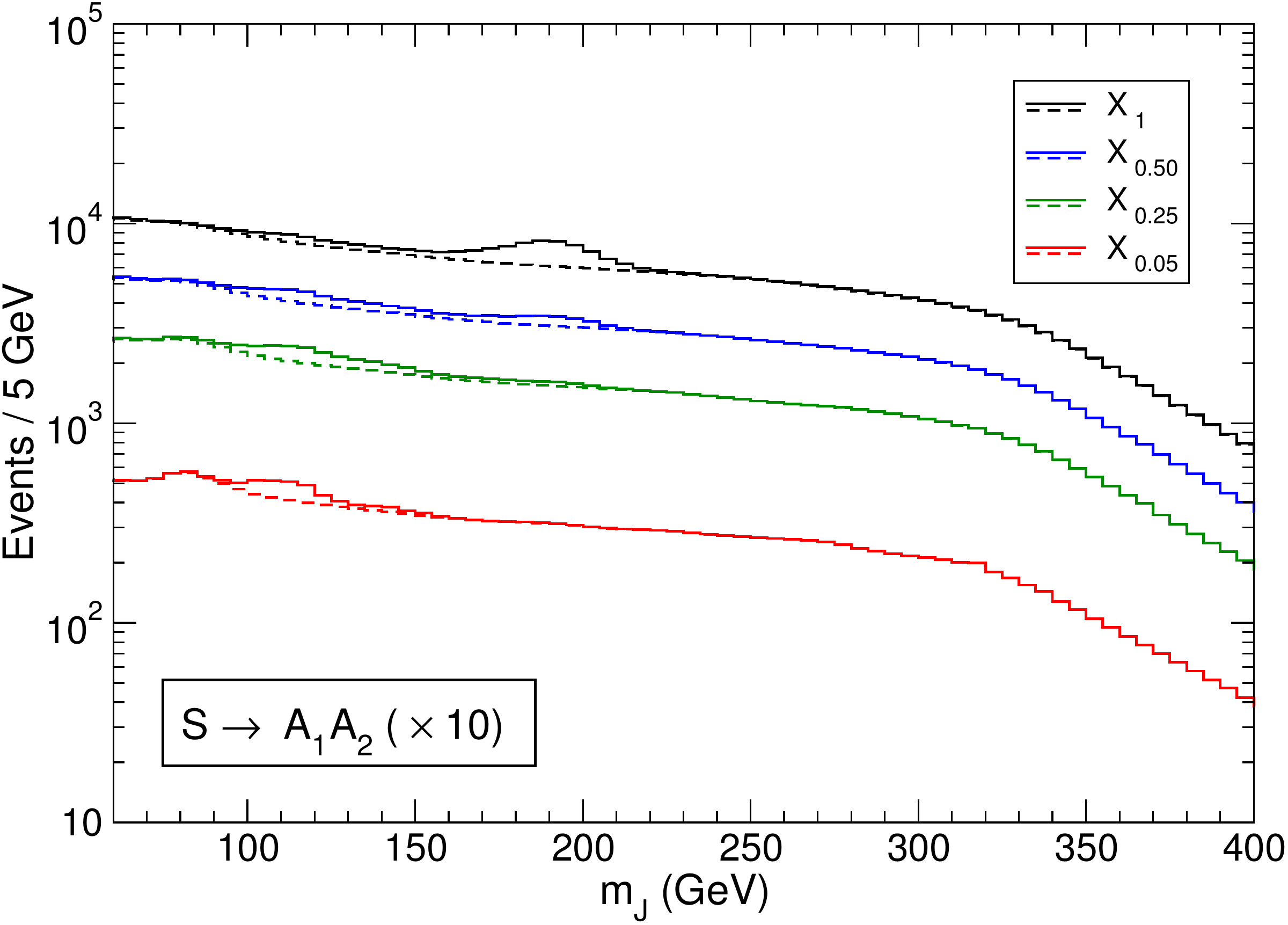} & 
\includegraphics[height=5.5cm,clip=]{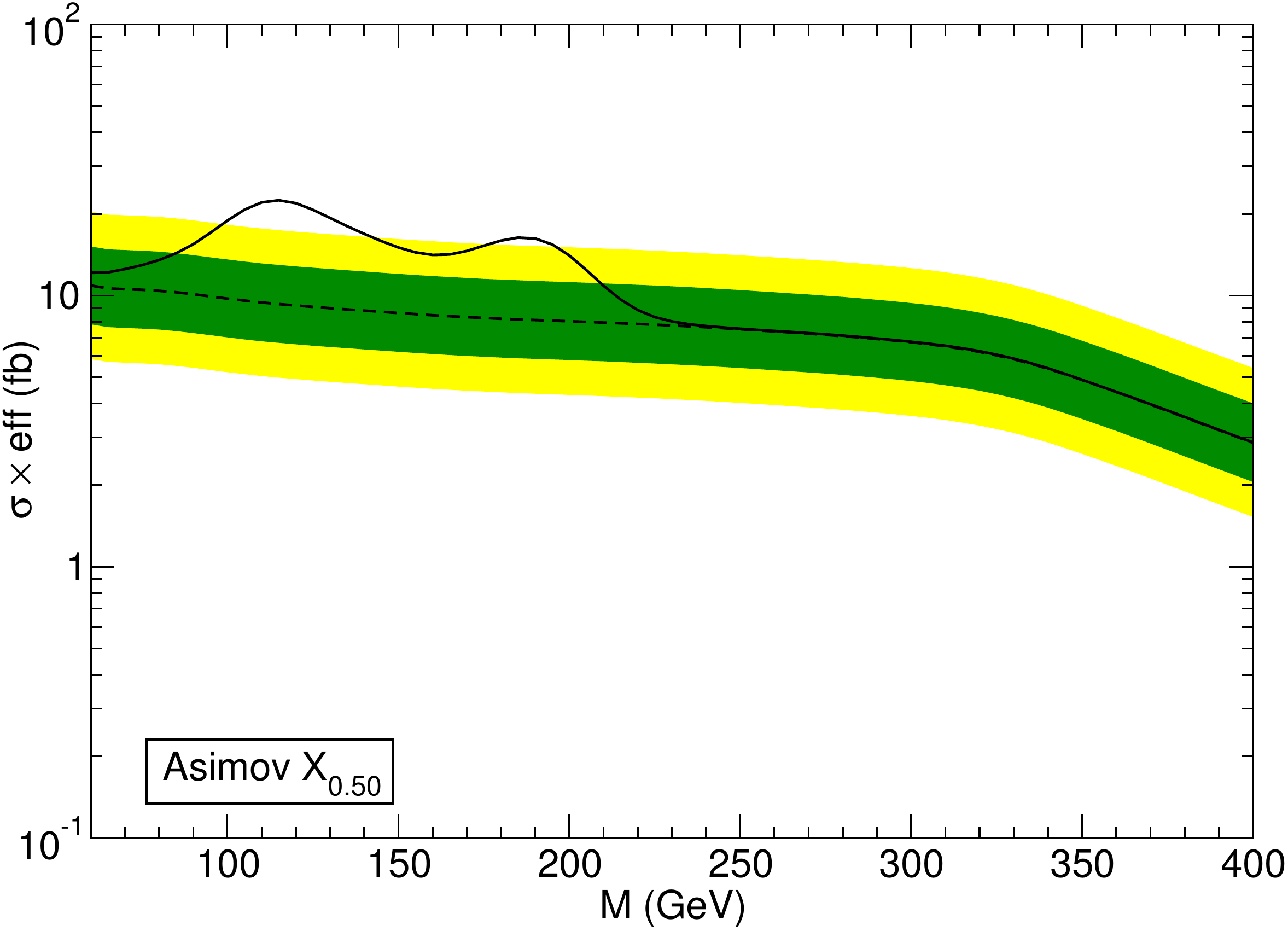} \\
\includegraphics[height=5.5cm,clip=]{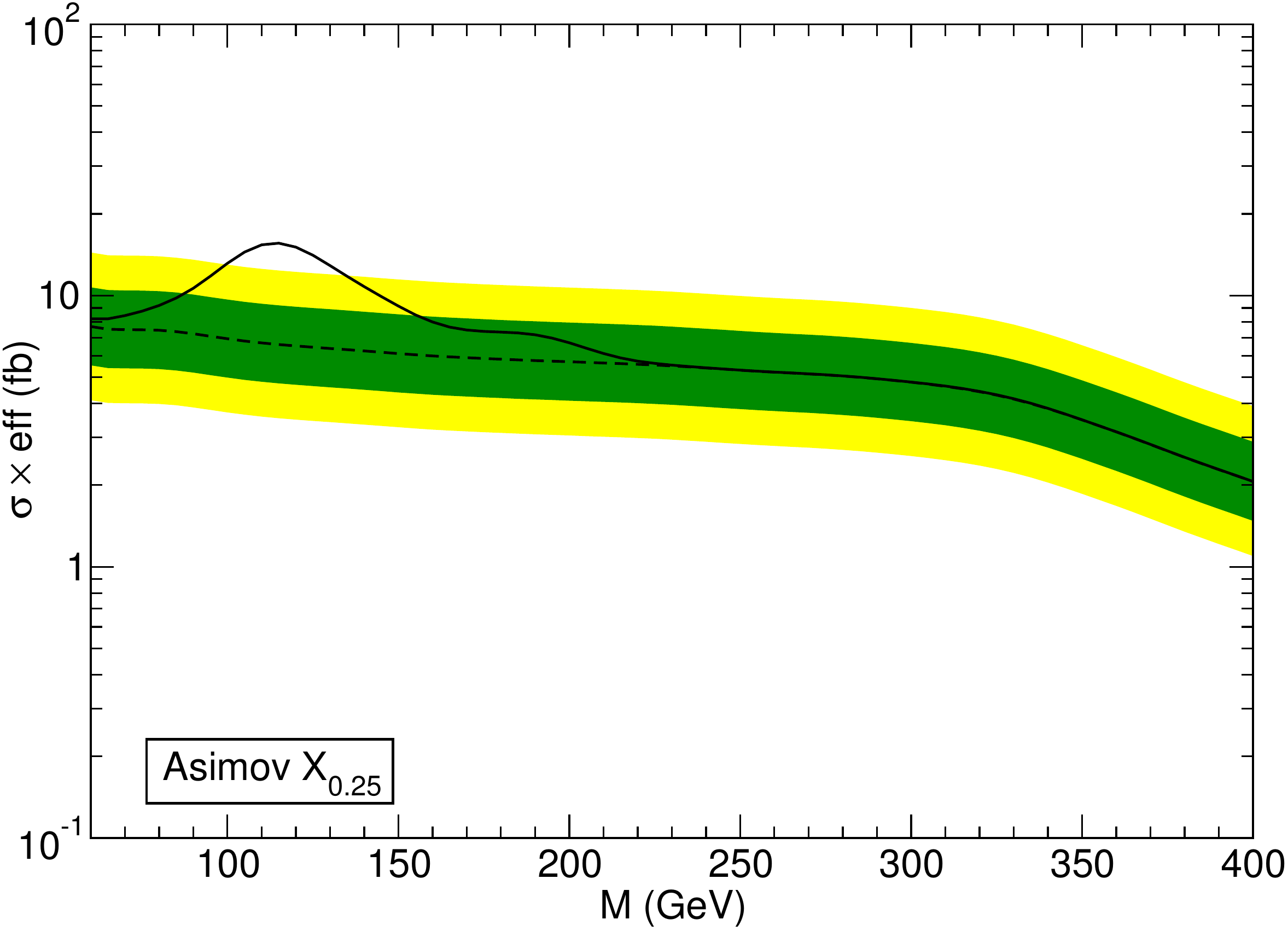}  &
\includegraphics[height=5.5cm,clip=]{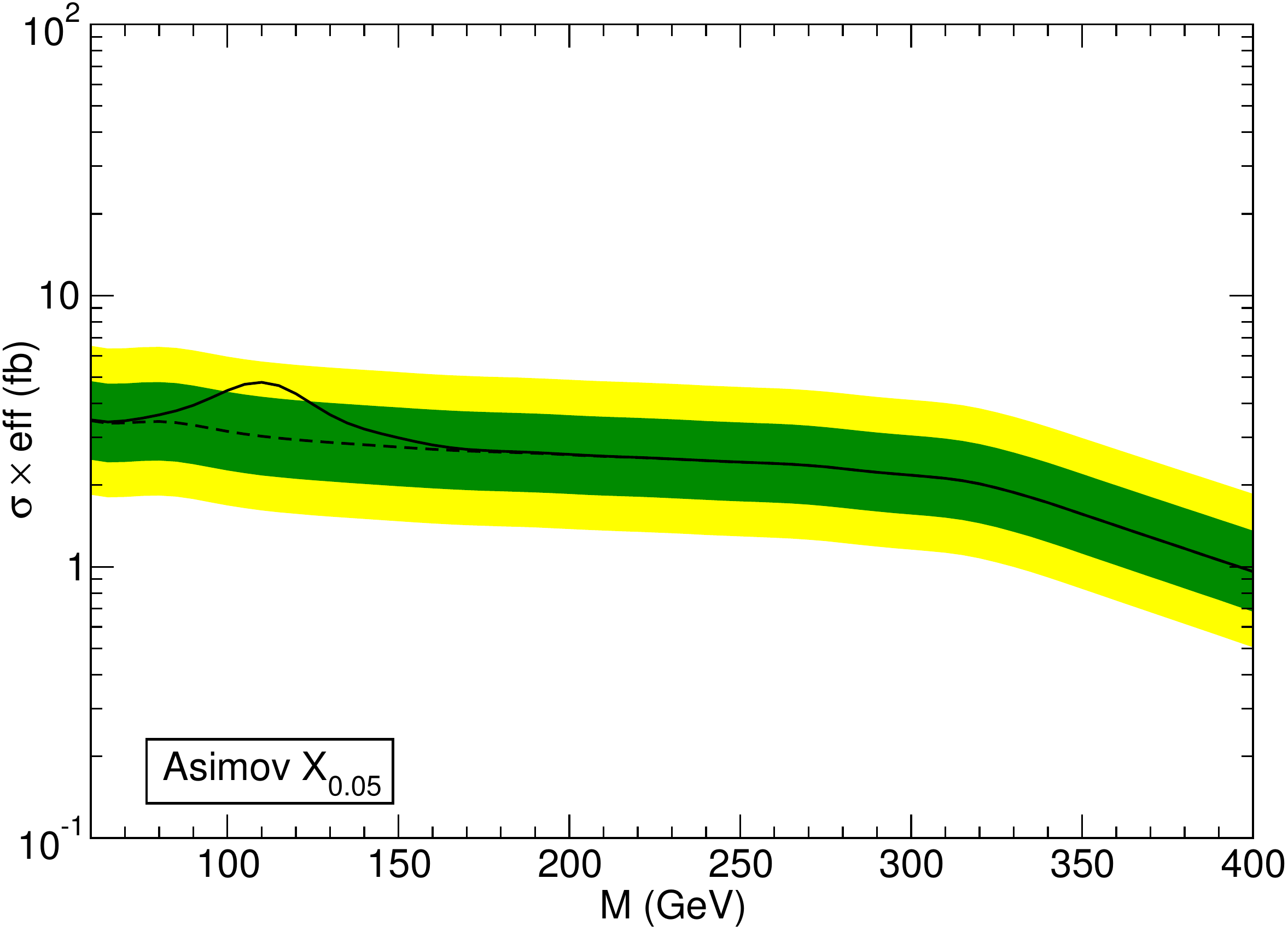}
\end{tabular}
\caption{The same as Fig.~\ref{fig:AA80}, for the $S \to A_1 A_2$ scenario.}
\label{fig:A115A20}
\end{center}
\end{figure}

The jet mass distributions at the different stages of the $N_2^1$ selection are presented for the three scenarios in the top, left panels of Figs.~\ref{fig:AA80}, \ref{fig:WW} and \ref{fig:A115A20}, respectively. The background plus injected signals correspond to the solid lines, while the dashed lines are the SM background. The small statistical fluctuations in the QCD background have been smoothed by a suitable algorithm that preserves the shape and the knee of the distribution. Also, for better visibility the size of the injected signals is multiplied by 10 in these plots. Without the $N_2^1$ cut, large and very wide bumps are observed at $M_S$ and below, in agreement with previous results~\cite{Aguilar-Saavedra:2017zuc}. These wide bumps may be difficult to detect because in this type of analyses, where the leading background is QCD multijet production, the background normalisation and the efficiency of the cut on $N_2^1$ or analogous jet substructure variables are usually calibrated from data (see also Ref.~\cite{Aguilar-Saavedra:2017iso}). Then, for example, a small modification of the shape of the knee near 300 GeV, as in Figs.~\ref{fig:AA80} and \ref{fig:WW}, may not be visible even if the extra number of events on the whole $m_J$ range is large --- remember that in these plots we have multiplied the signal by a factor of 10. With the tighter selection on $N_2^1$ the bump at $M_S$ slightly moves to lower masses and the secondary bump at $M_{A,W} = 80$ GeV (for the first and second scenarios) and  $M_{A_2} =115$ GeV (for the third scenario) becomes more prominent. In this latter scenario, a third smaller bump appears at $M_{A_1} = 20$ GeV too, but it is removed by the cut on $\rho$.

It is also interesting to consider how these bumps would show up in the observed limits on new physics signals. With this purpose, we perform  likelihood tests for the presence of narrow resonances over the expected background, using
the $\text{CL}_\text{s}$ method~\cite{Read:2002hq} with the asymptotic approximation of Ref.~\cite{Cowan:2010js}.
We use for pseudo-experiments the Asimov dataset including the injected signals,\footnote{Following Ref.~\cite{Cowan:2010js} we denote the Asimov dataset as the one where the observed data correspond to the mean of the corresponding distributions, bin by bin.} in order to isolate the effect discussed from statistical fluctuations. The probability density functions of the potential narrow resonance signals are Gaussians with centre $M$ (i.e. the resonance mass probed) and standard deviation of 10 GeV. We do not include any systematic uncertainty in the form of nuisance parameters, as these do not affect our arguments, only decreasing the statistical significance of the bumps.

The 95\% confidence level (CL) upper limits on cross section times efficiency, for the $X_{0.50}$, $X_{0.25}$ and $X_{0.05}$ working points, are collected in Figs.~\ref{fig:AA80} (for $S \to AA$), \ref{fig:WW} ($S \to WW$) and \ref{fig:A115A20} ($S \to A_1 A_2$). The trend is the same in the three scenarios considered, with small differences in the relative size of the high- and low-mass bumps, and follow what one expects from Fig.~\ref{fig:avgN2} and the above discussion:
\begin{enumerate}
\item[(i)] with a looser $N_2^1$ selection the high-mass bump has a larger statistical significance than the low-mass one;
\item[(ii)] a more stringent $N_2^1$ selection wipes out the high-mass bump but may enhance the significance of the low-mass one;
\item[(iii)] an even more stringent $N_2^1$ selection ends up reducing the significance of the low-mass bump as well.
\end{enumerate}
The bump running effect has two ingredients: first, the appearance of a secondary mass peak away from $M_S$ due to the jet grooming; second, the suppression of the large mass bump near $M_S$ by a tight selection on $N_2^1$. With a loose selection on $N_2^1$, both bumps coexist and the high mass bump slightly moves towards lower masses.
As we show in appendix~\ref{sec:a}, this effect has little dependence on the transverse momentum of the stealth boson signals. And it happens to varying degrees when the jets are groomed using the trimming~\cite{Krohn:2009th} or pruning~\cite{Ellis:2009me} algorithms, and also for larger jet radii, as seen in Ref.~\cite{Aguilar-Saavedra:2017zuc}. A less aggresive jet grooming, as investigated in appendix~\ref{sec:b}, decreases the size of the secondary mass peak; however, it it not clear whether a milder grooming may provide an adequate jet mass resolution in an intense pile-up environment such as the LHC Run 2.

The obvious consequence of the bump running effect is that one may see an excess at a given mass, say $M_W$, and interpret that this is due to the production of a $W$ boson, while it is actually due to a new, much heavier particle. And, while for actual $W$ and $Z$ bosons one expects signals in the leptonic channels, for these stealth boson signals the leptonic modes may be absent. For example, in $S \to WW$ the semileptonic decay of the $WW$ pair gives rise to a fat jet from one $W$ which contains a very energetic lepton from the other boson; this kind of signature has not been experimentally searched for, to our knowledge. The leptonic decay of the $WW$ pair gives rise to two collimated leptons plus missing energy, which is not the standard signature from a leptonic $W$ decay.

\section{Other related effects}
\label{sec:4}

The bump running effect and the appearance of double (or triple) bumps may lead to some other puzzling effects when comparing different analyses, i.e. different event selections, or different kinematical regions in standard searches for simple topologies. We discuss here two of particular relevance for the interpretation of current searches using simplified models as benchmarks.

\subsection{Fake flatness}
\label{sec:4.1}

\begin{figure}[t]
\begin{center}
\includegraphics[height=5.5cm,clip=]{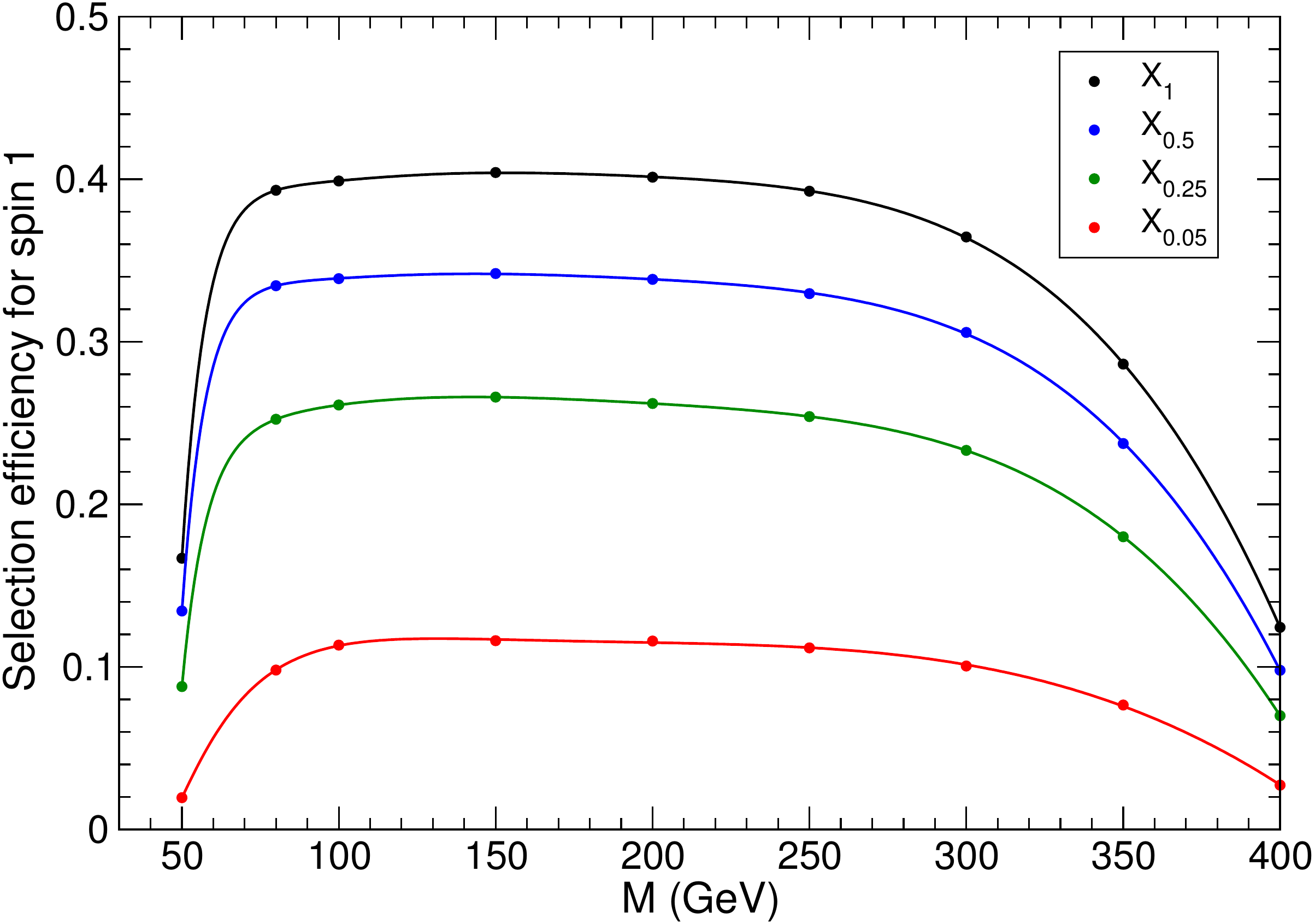} 
\caption{Selection efficiency for the decay of a heavy resonance into a light vector boson, as a function of its mass.}
\label{fig:eff}
\end{center}
\end{figure}

\begin{figure}[t]
\begin{center}
\begin{tabular}{cc}
\includegraphics[height=5.4cm,clip=]{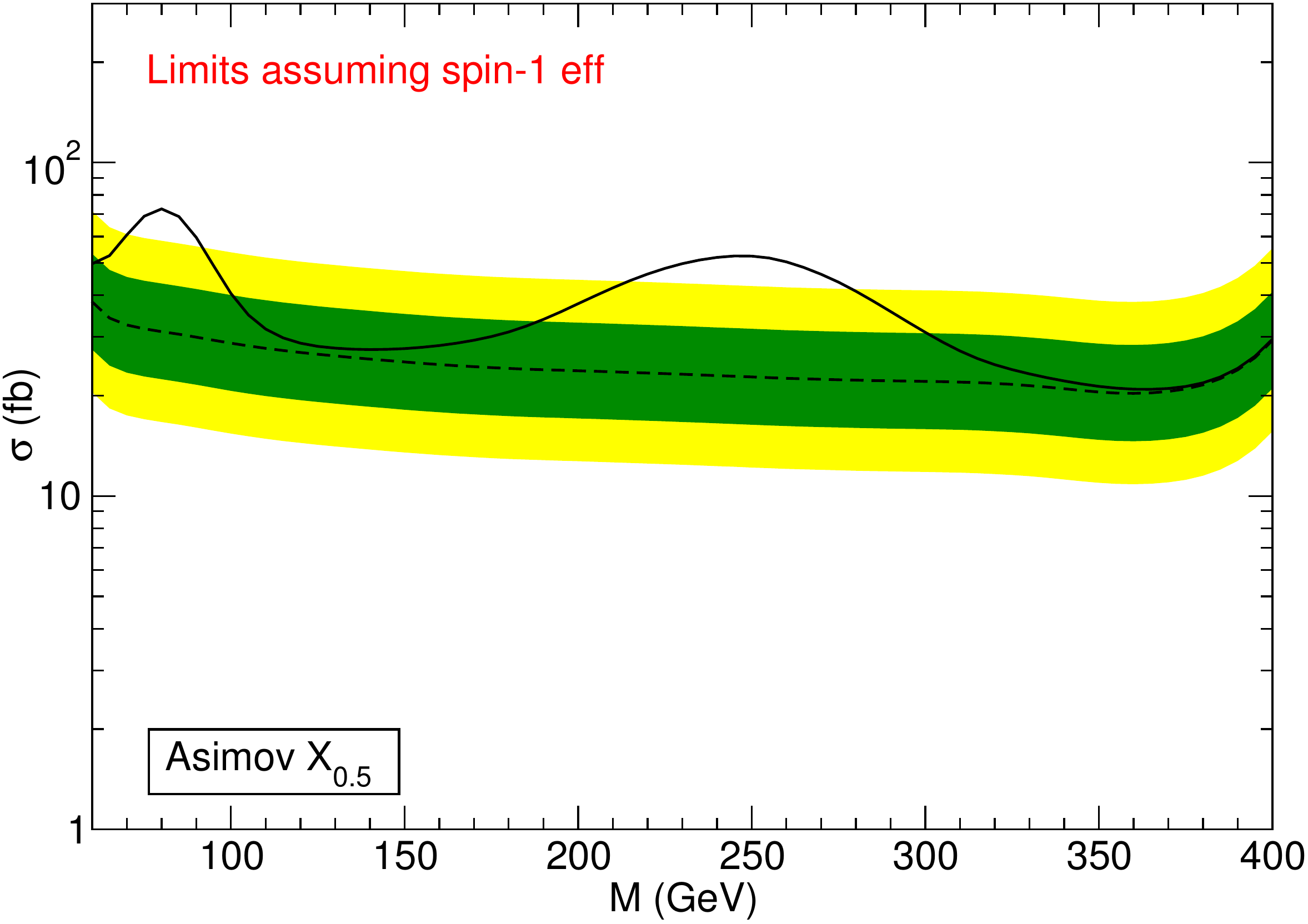} & 
\includegraphics[height=5.4cm,clip=]{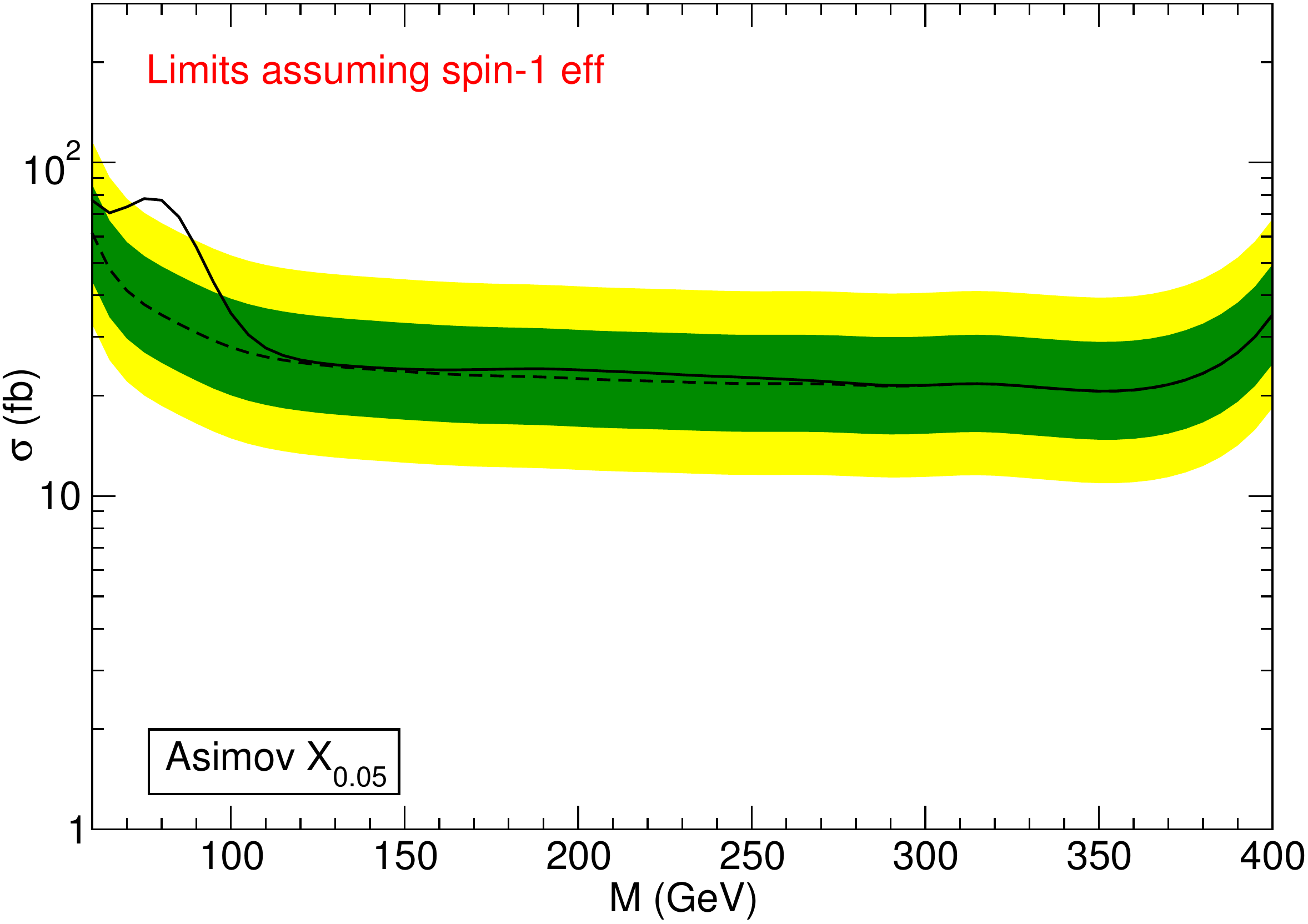} 
\end{tabular}
\caption{Expected and observed limits on narrow resonances corresponding to the $WW$ signal injected, interpreted as limits on the production of a light vector boson.}
\label{fig:limits}
\end{center}
\end{figure}

The upper limits shown in Figs.~\ref{fig:AA80}--\ref{fig:A115A20} are placed on signal cross section times selection efficiency. Usually, experiments report the results of the searches in terms of cross sections for a given model with a certain efficiency. Model interpretations must always be taken with a grain of salt; still, it is illuminating to consider what might happen when `interpreting' the bumps arising from stealth bosons within a simple model. We can for example translate the expected and observed limits for the $S \to WW$ scenario in Fig.~\ref{fig:WW} by assuming the selection efficiency for a light vector boson $W' / Z' \to q \bar q$ (instead of the actual efficiency for $S \to WW \to q \bar q q \bar q$) resulting from the decay of a heavy resonance, that is, replacing $H_1^0$ in Eqs.~(\ref{ec:signals}) by a vector boson. The selection efficiencies computed for different masses are shown in Fig.~\ref{fig:eff}, together with a smooth interpolation. The result of this light vector boson interpretation of the limits is shown in Fig.~\ref{fig:limits}, for the $X_{50}$ (left) and $X_{0.05}$ (right) working points. While the low-mass bumps at $M_W$ are compatible, the large bump at 250 GeV on the left plot is in clear tension with the null result on the right plot. Should two experiments present these two results, one would easily conclude that the bump on the left plot is a statistical fluctuation, excluded by the right plot, when it is actually the model interpretation that is biasing the comparison.

\subsection{Sideband contamination}

Let us consider the decay of a heavy resonance into a stealth boson and a weak boson, taking for definiteness $Z' \to H_1^0 Z$ as in Eqs.~(\ref{ec:signals}), with the $Z$ boson decaying leptonically and the stealth boson $S = H_1^0$ giving a fat jet. When the groomed jet mass happens to be close to $M_{W,Z}$ the signal is diboson-like, and can be detected by standard diboson searches in the semileptonic $\ell \ell J$ channel~\cite{Aaboud:2017itg,CMS:2017vix}, with $\ell$ a charged lepton (electron or muon). These searches address final states with two charged leptons with invariant mass consistent with $M_Z$, and a jet with groomed mass in the $M_{W,Z}$ range, subject to some loose tagging requirement using $\tau_{21}$ (CMS) or $D_2$ (ATLAS). For example, the CMS analysis in Ref.~\cite{CMS:2017vix} uses a signal region with jet mass $m_J \in [65,105]$ GeV. For background normalisation, these analyses use sideband regions with $m_J$ outside the signal region. In the case of stealth bosons, an important sideband contamination can be produced by the high-mass bump around $M_S$. This potential contamination does not strongly depend on whether the jet substructure variables are measured on groomed or ungroomed jets, and it may happen in the low-mass sideband too if one of the $S$ decay products is lighter.

In order to assess the size of this contamination, we use a generic event selection similar to the ones used by the ATLAS and CMS Collaborations. We consider events having two charged leptons with $p_T > 40$ GeV, and pseudorapidity $|\eta| < 2.5$ for electrons and $|\eta| < 2.4$ for muons. Their invariant mass must lie in the range $60 < m_{\ell \ell} < 120$ GeV. The same criteria applied to jets in section~\ref{sec:2} are used, defining $m_J \in [65,105]$ as the signal region, and a high-mass sideband $m_J > 105$ GeV. The signals considered are those in Eqs.~(\ref{ec:signals}) but with leptonic decay of the $Z$ boson.

\begin{figure}[t]
\begin{center}
\begin{tabular}{cc}
\includegraphics[height=5.3cm,clip=]{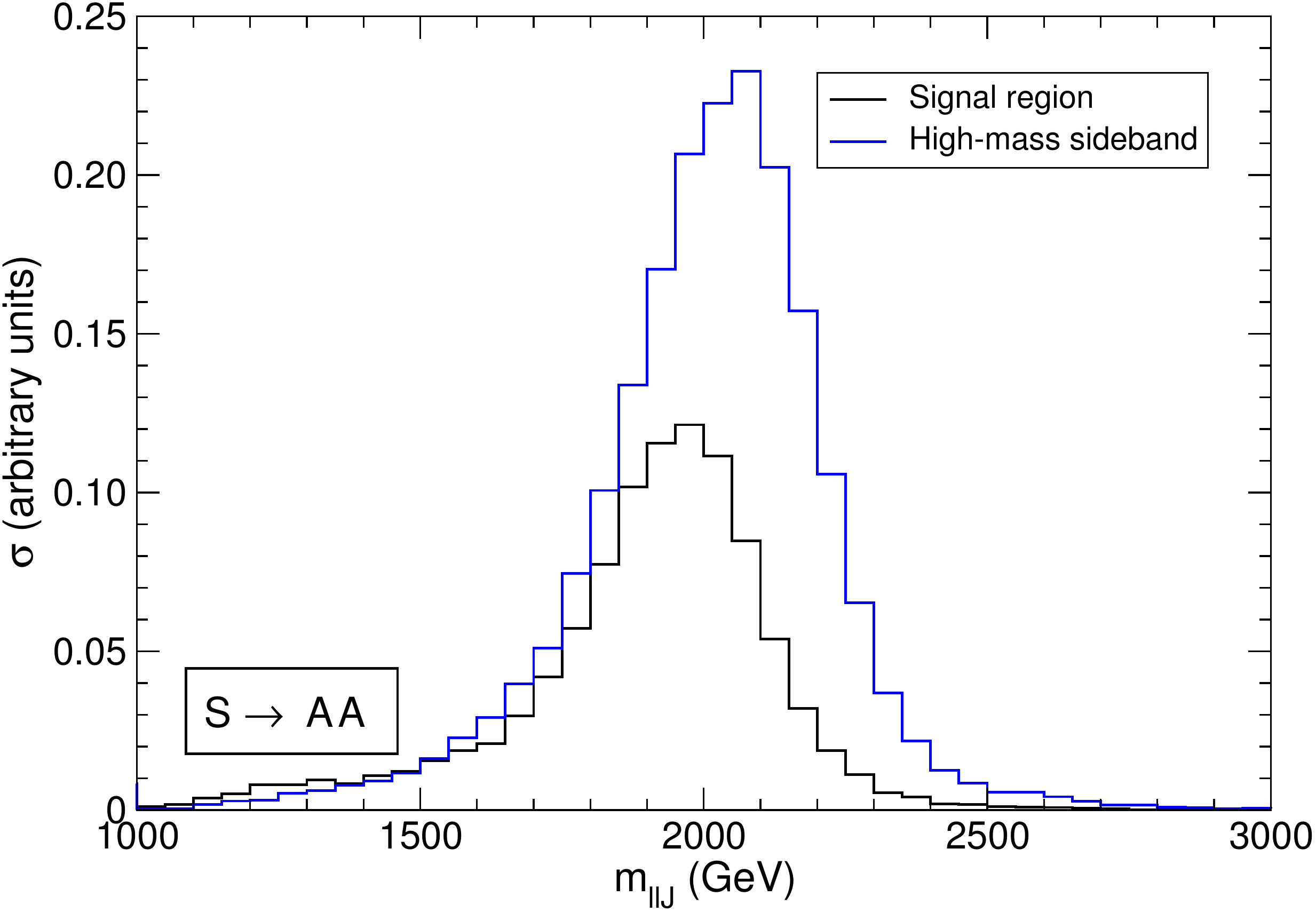} & 
\includegraphics[height=5.3cm,clip=]{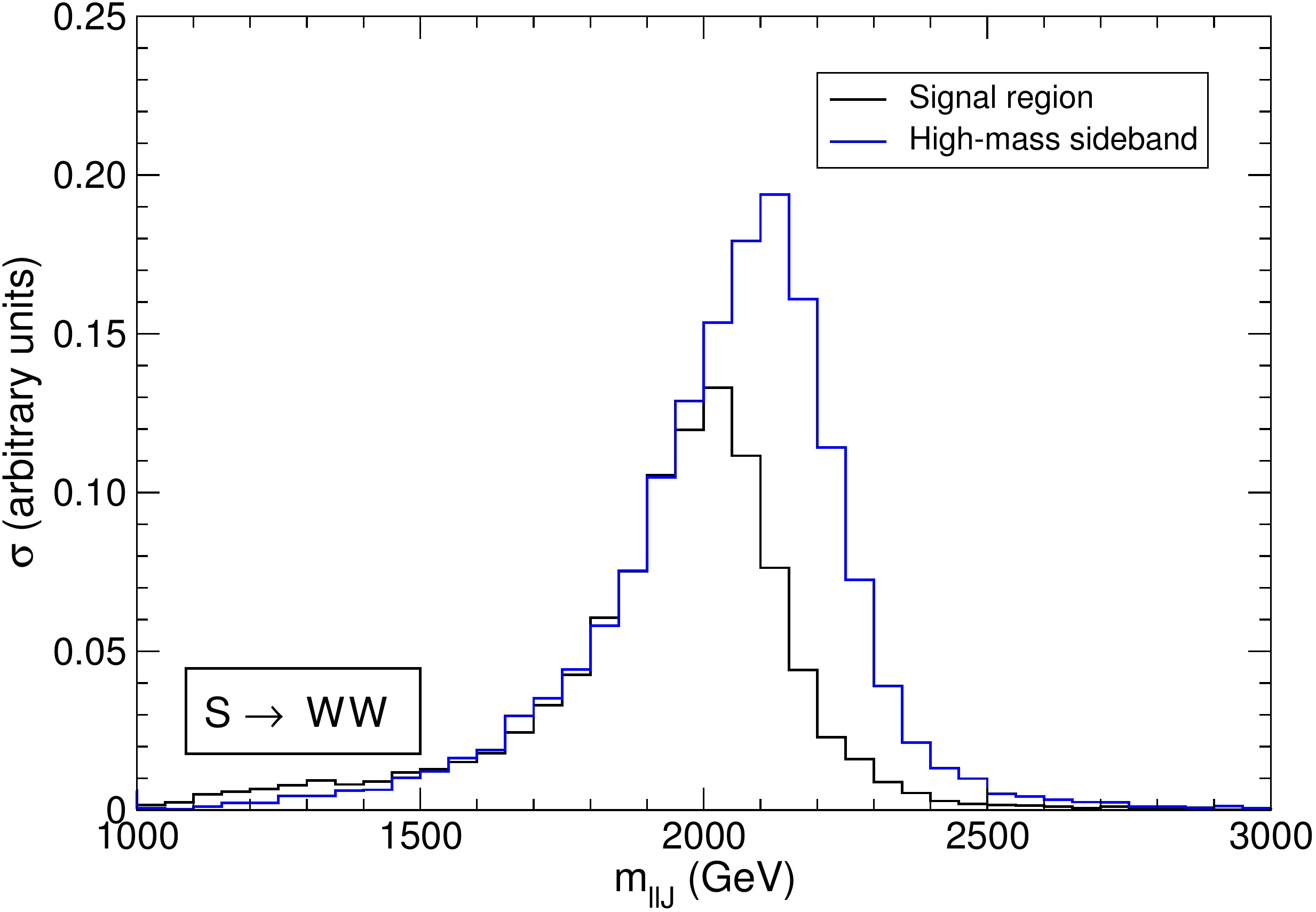} 
\end{tabular}
\caption{$\ell \ell J$ invariant mass distribution for events in the signal region (black) and the high-mass sideband (blue), for the $S \to AA$ scenario (left) and $S \to WW$ (right).}
\label{fig:sideband}
\end{center}
\end{figure}

The $\ell \ell J$ invariant mass distribution, which is a proxy for the heavy resonance mass, is plotted in Fig.~\ref{fig:sideband} for events in the signal region and in the high-mass sideband, for $S \to AA$ (left panel) and $S \to WW$ (right panel). The centre of the distribution is shifted between the signal region and high-mass sideband, an obvious consequence of the difference in the jet mass.
In this example, the sideband contribution is twice larger than in the signal region, but the relative size can even increase, depending on several factors, for example the jet tagging working point and the jet transverse momentum. In any case, it is clear that this type of signals can dangerously pollute the control regions of standard diboson searches.

\section{Discussion}
\label{sec:5}

The first, striking consequence of the bump running effect discussed in this paper is that a stealth boson can appear to have the mass of one of its decay products. And this identity confusion would lead to a puzzling behaviour. For example, should we observe in a search a bump involving a jet mass $m_J \sim M_W$ (caused by the jet grooming) and no other bump (as consequence of the jet substructure cut), we would arguably consider that we are dealing with a hadronically-decaying $W$ boson, and look for companion signals when the $W$ boson decays leptonically. But those signals would not be present. (The same can happen with the $Z$ boson, for stealth boson decays $S \to ZA$ or $S \to ZZ$.)
And, unless the statistical significance of the bump were in excess of $5\sigma$ ---which is basically impossible to achieve for such an elusive signal without a dedicated analysis--- we would catalog the bump as a mere fluctuation or systematic effect.
Previous literature~\cite{Aguilar-Saavedra:2015rna,Aguilar-Saavedra:2016xuc} has also addressed these apparent inconsistencies, where a triboson resonance signal might be seen in the diboson resonance searches in hadronic channels~\cite{Aad:2015owa,Khachatryan:2014hpa,TheATLAScollaboration:2015msj,ATLAS:2016yqq,Sirunyan:2017acf} but not in the leptonic ones. We point out that more of such anomalies exist, for example a CMS search for $Z\gamma$ resonances~\cite{Sirunyan:2017hsb} finds 
a $3.2\sigma$ broad excess at 2 TeV in the $Z \to q \bar q$ hadronic channel, without a counterpart in the leptonic channels.

For this effect to be attenuated, a grooming algorithm that is more robust for multi-pronged jets is highly desirable. We have investigated in appendix~\ref{sec:b} how the size of the secondary low-mass bump decreases with a less aggresive grooming. The results are not completely satisfactory, as the bump does not disappear for moderate variations from the `reference' parameters used by the ATLAS and CMS Collaborations, for which the soft drop algorithm is found to perform well under the intense pile-up conditions at Run 2. And, at the same time, the resolution of the high-mass bump slightly decreases with the change of parameters.

Independently of the above, jet substructure variables computed from the ungroomed jet, as used by the CMS Collaboration in most analyses~\cite{CMS:2017vix,Sirunyan:2017acf,CMS:2017vdr}, are preferred, as they are not influenced by a possible bias from the grooming. In particular, a generic anti-QCD tagger~\cite{Aguilar-Saavedra:2017rzt} that does not penalise multi-pronged signals always constitutes an advantage when looking for signals yielding non-standard jets.

Model-dependent interpretations can be very misleading, as it is well known, and we have seen here an example: when considering two different signal regions, corresponding to two choices for the $N_2^1$ thresholds, we can obtain apparently contradictory results: with the looser selection a large high-mass bump is present, which is almost excluded at the 95\% CL by the tighter selection. This reminds us that, when comparing the results of two or more experiments, the underlying assumptions used to present the results have to be carefully taken into account.

Finally, we have seen that stealth bosons giving multiple mass bumps can simultaneously contribute to signal regions and sidebands in standard searches. This is a `nightmare scenario' that can be attenuated with model-independent tools~\cite{Aguilar-Saavedra:2017rzt}, or avoided by dedicated searches. In this context, it is worthwhile noting that the ATLAS diboson resonance search in the $\ell \ell J$ channel~\cite{Aaboud:2017itg} observes a $\sim 3\sigma$ dip at $M=800$ GeV, a similar dip is seen by the CMS Collaboration in the same channel~\cite{CMS:2017vix} around $M = 750$ GeV, and a $\sim 2\sigma$ dip is seen at $M=800$ GeV in an ATLAS search for $ZH$ resonances in the $\ell \ell J$ channel~\cite{Aaboud:2017cxo}. While it is premature to make any claim, especially without a detailed recast of these searches, the previous experience with the CDF $Wjj$ excess~\cite{Aaltonen:2011mk} shows that an incorrect background normalisation can fake narrow `signal' peaks. These underfluctuations and the possibility of a mismodeling deserve further investigation.

\section*{Acknowledgements}
This work has been supported by MINECO Projects  FPA 2016-78220-C3-1-P and FPA 2013-47836-C3-2-P (including ERDF) and by Junta de Andaluc\'{\i}a Project FQM-101.

\appendix

\section{Groomed versus ungroomed jets}
\label{sec:a}

\begin{figure}[htb]
\begin{center}
\begin{tabular}{ccc}
\includegraphics[height=4cm,clip=]{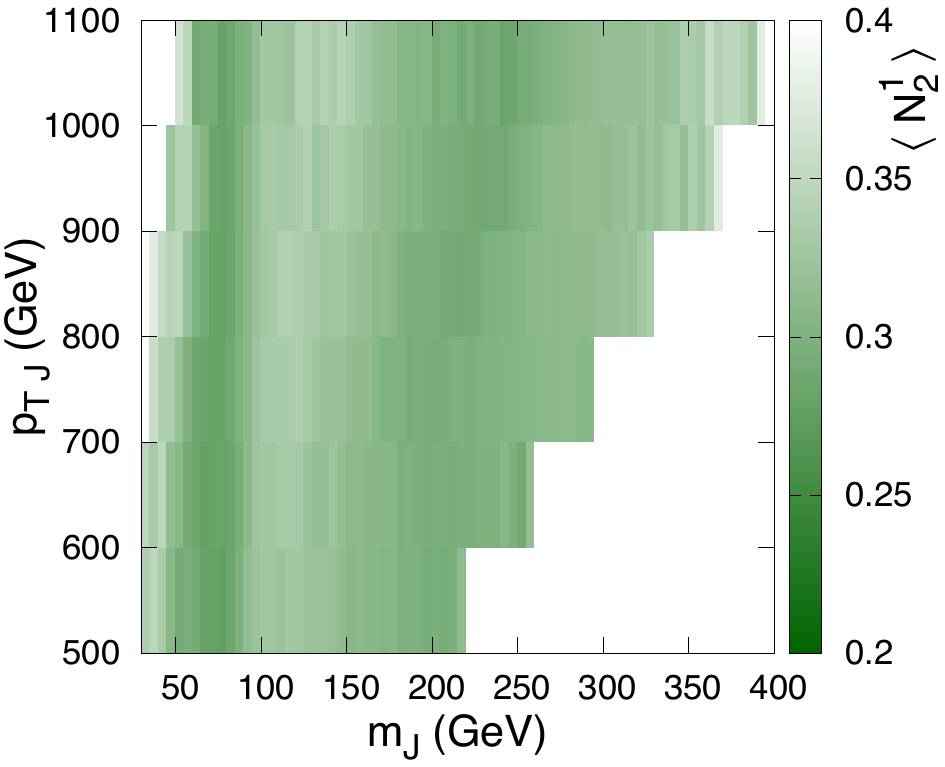} & 
\includegraphics[height=4cm,clip=]{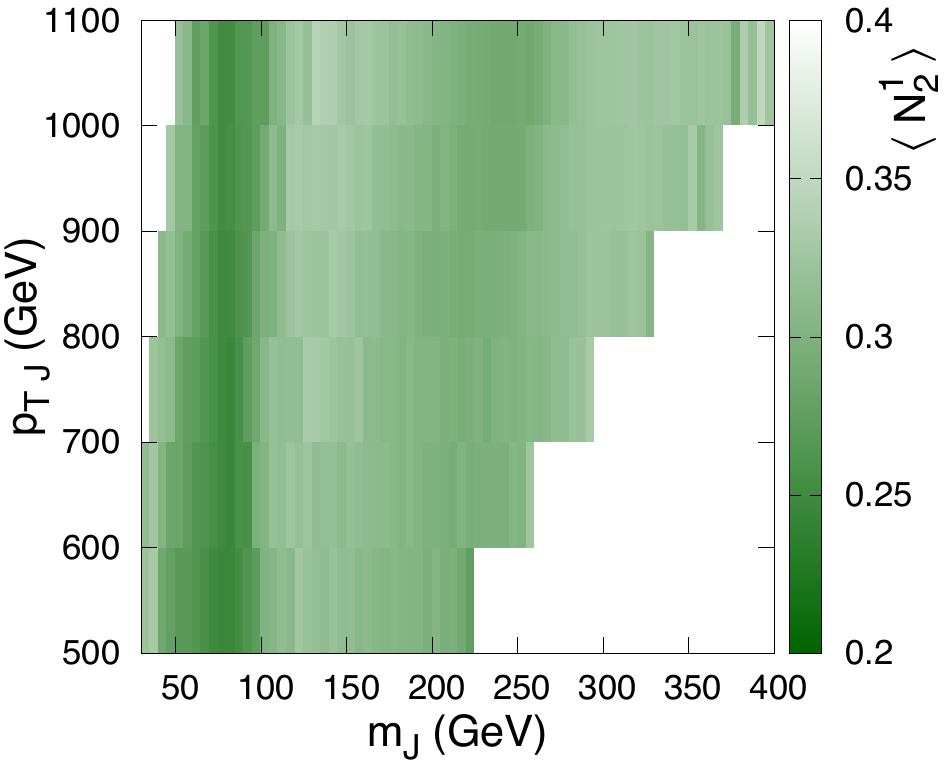} &
\includegraphics[height=4cm,clip=]{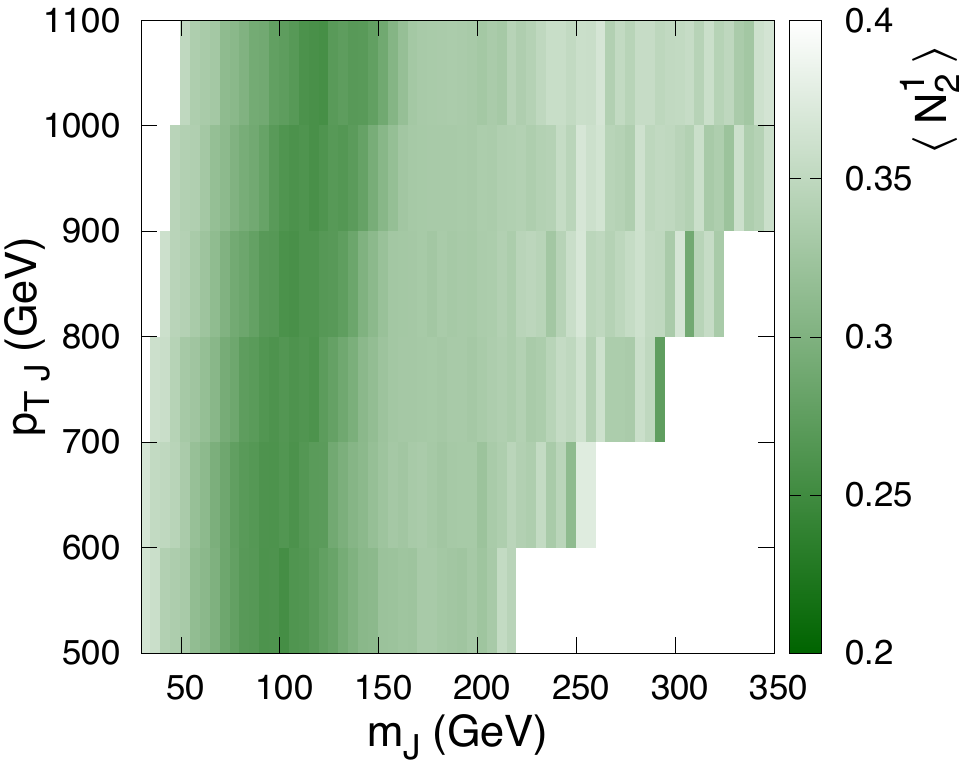}  \\
\includegraphics[height=4cm,clip=]{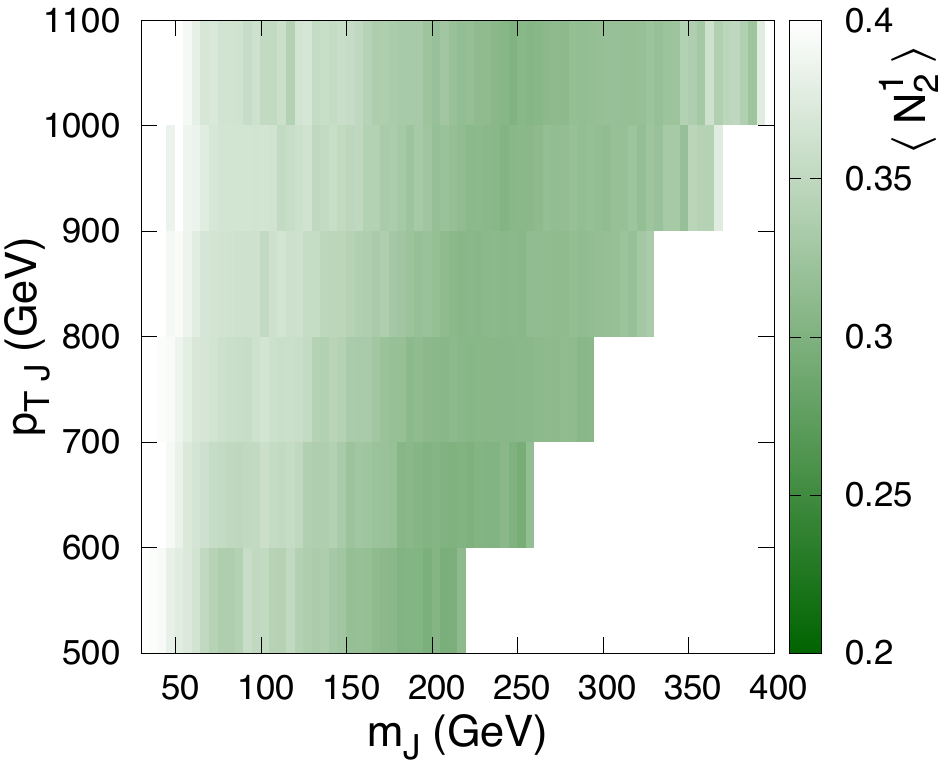} & 
\includegraphics[height=4cm,clip=]{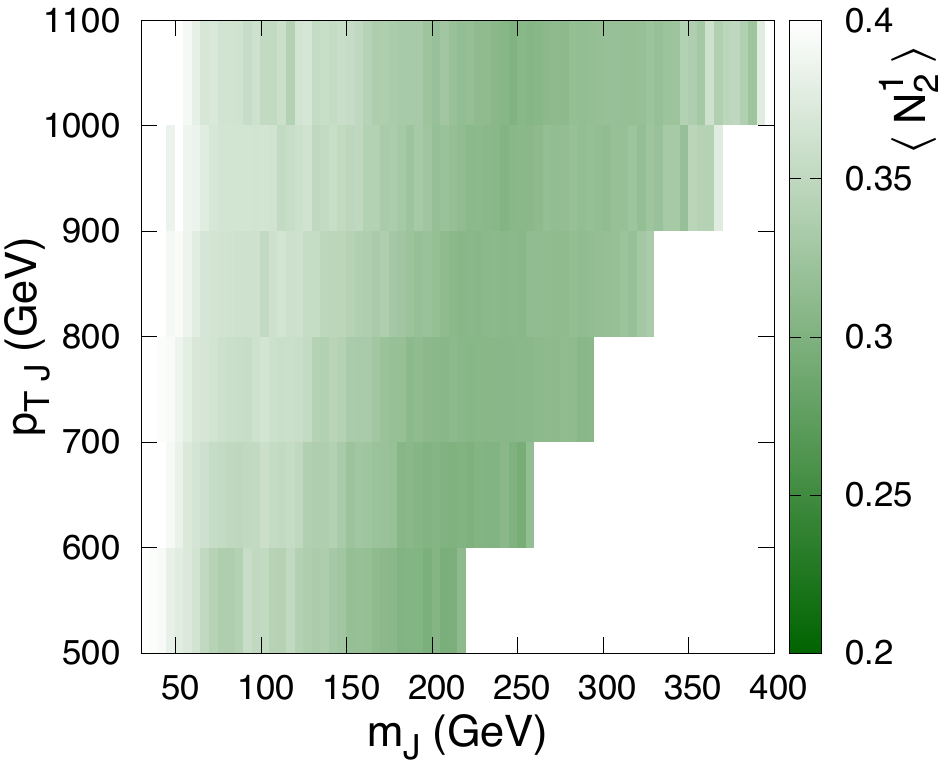} &
\includegraphics[height=4cm,clip=]{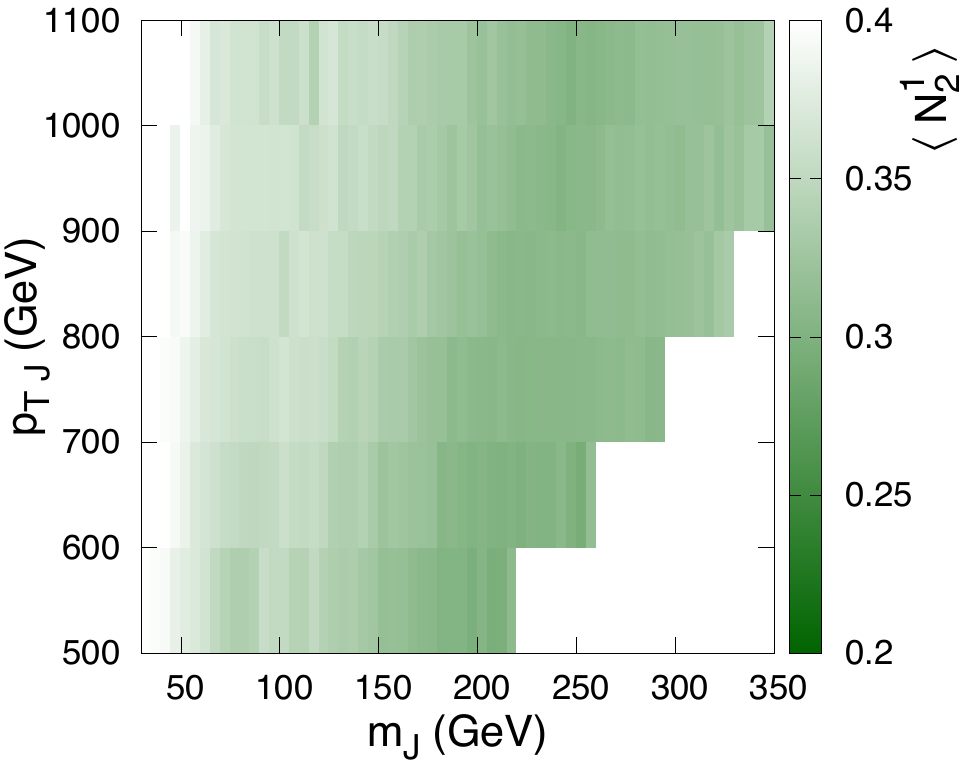} 
\end{tabular}
\caption{Dependence on the jet mass and transverse momentum of the average $\langle N_2^1 \rangle$ for groomed jets (top) and ungroomed jets (bottom). The left, middle and right panels correspond to $S \to AA$, $S \to WW$ and $S \to A_1 A_2$, respectively. The white areas at the lower right corner are not populated. }
\label{fig:avgN2comp1}
\end{center}
\end{figure}

In addition to jet mass, jet substructure variables such as $N_2^1$ depend on the jet transverse momentum. We investigate that dependence for stealth boson signals by plotting the average $\langle N_2^1 \rangle$ in Fig.~\ref{fig:avgN2comp1}, for groomed and ungroomed jets, and for the three stealth boson scenarios. In all cases, $m_J$ and $\ptj$ correspond to the groomed quantities. For groomed jets we can see that the `dips' in the  $\langle N_2^1 \rangle$ versus mass distribution have a very mild dependence on $\ptj$. For ungroomed jets, the dependence on $\ptj$ is very weak, too.
\begin{figure}[t]
\begin{center}
\includegraphics[height=5.5cm,clip=]{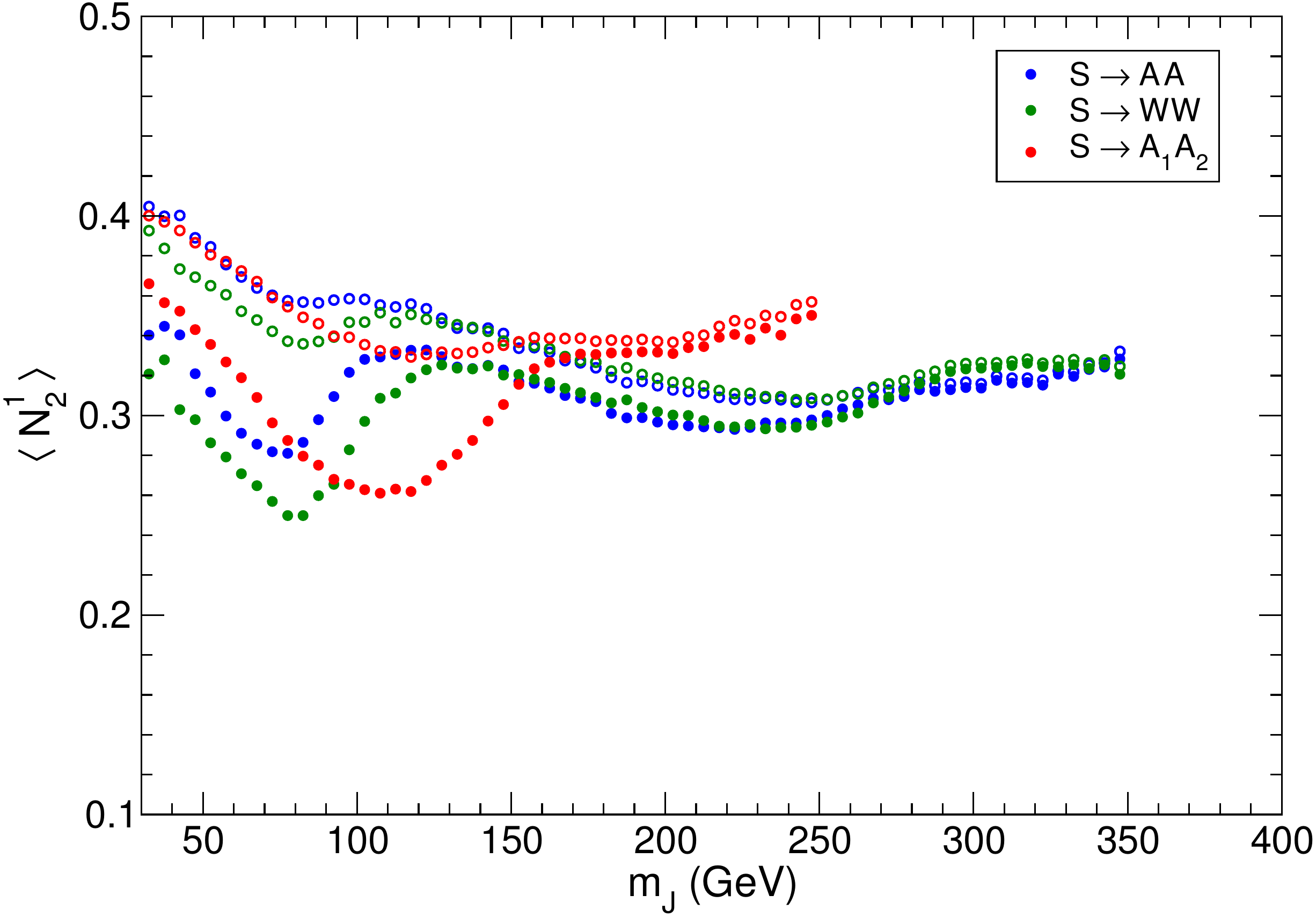} 
\caption{Dependence on the jet mass of the average $\langle N_2^1 \rangle$, for groomed jets (solid circles) and ungroomed jets (hollow circles).}
\label{fig:avgN2comp2}
\end{center}
\end{figure}
In Fig.~\ref{fig:avgN2comp2} we show the comparison between $\langle N_2^1 \rangle$ for groomed and ungroomed jets, integrated for all $\ptj$ range. (In contrast with Fig.~\ref{fig:avgN2}, we do not consider the decorrelated quantities by subtracting $X_{0.50}$ because the latter differs in the two cases.)
As previously indicated, the dips at the mass of the secondary resonance, $M_A$, $M_W$ or $M_{A_2}$, are not present for ungroomed jets. This plot also shows that the presence of the dips is not a consequence of the decorrelation between jet mass and tagging efficiency, and therefore it is also expected when other decorrelation procedures~\cite{Dolen:2016kst,Moult:2017okx} are used.

For completeness, let us also discuss the interplay between grooming and jet substructure when used mixed groomed/ungroomed jet subjettiness variables. The subjettiness ratio $\tau_{21} = \tau_2 / \tau_1$~\cite{Thaler:2010tr} is often used to tag weak bosons decaying hadronically, in most analyses using the ungroomed jets~\cite{CMS:2017vix,Sirunyan:2017acf,CMS:2017vdr} but sometimes it is also used on the groomed jets~\cite{Aaboud:2018zba}. A different proposal~\cite{Salam:2016yht} advocates for the use of the so-called `dichroic' ratios, with $\tau_2$ computed for the ungroomed jet and $\tau_1$ for the groomed jet. In light of the foregoing arguments, one expects that when the grooming removes one of the stealth boson decay products, (i.e. when the mass bump is shifted to lower values) the value of $\tau_1$ will drastically decrease, so the mixed ratio $\tau_{21}^{(\text{dic})} = \tau_2^{(\text{ung})} / \tau_1^{(\text{gr})}$ will be enhanced, opposite to what happens when groomed variables are used everywhere.

\begin{figure}[htb]
\begin{center}
\includegraphics[height=5.5cm,clip=]{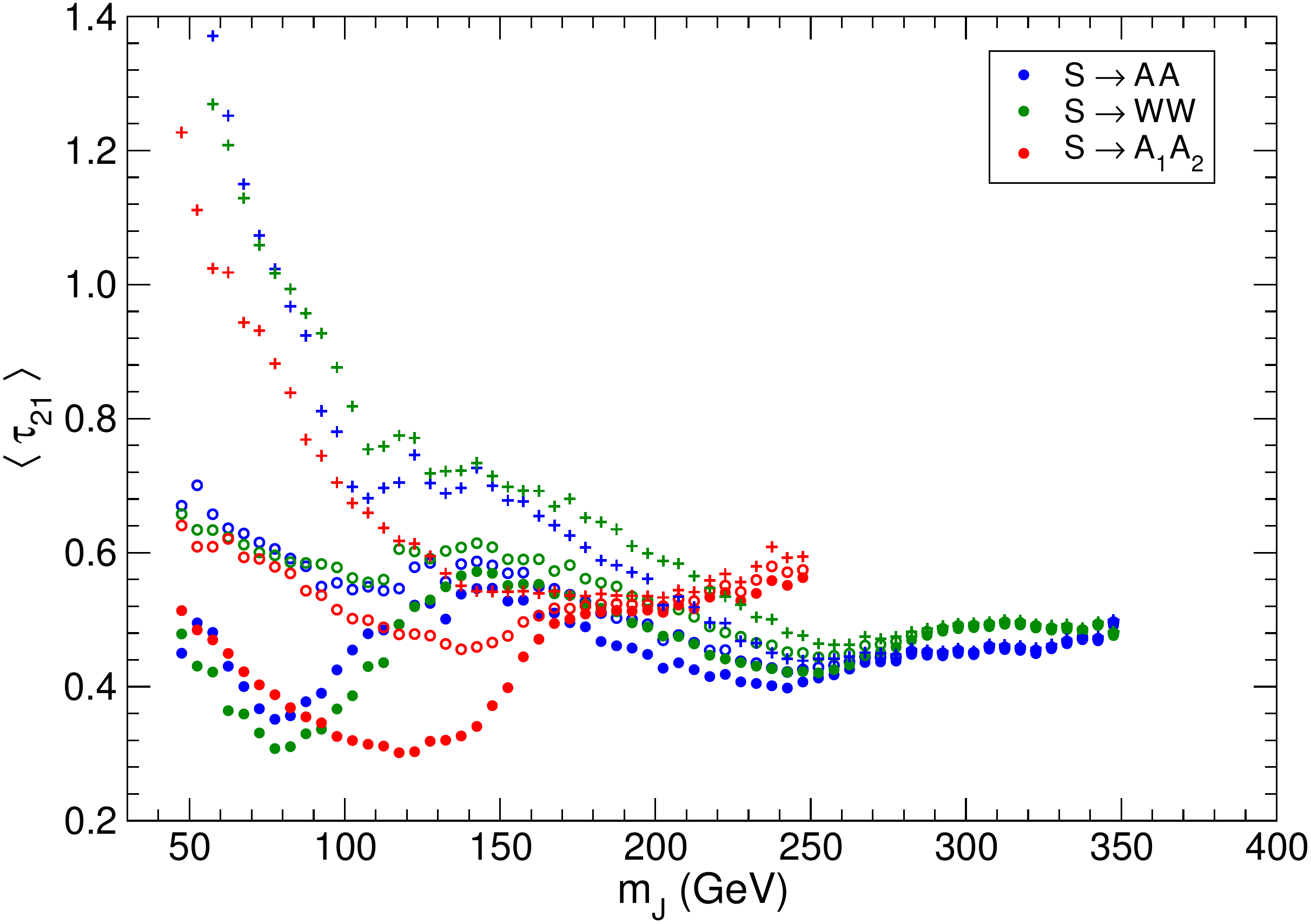} 
\caption{Dependence on the jet mass of the average $\langle N_2^1 \rangle$, for groomed jets (solid circles) and ungroomed jets (hollow circles).}
\label{fig:avgtau2comp}
\end{center}
\end{figure}

We can observe this behaviour by computing the average $\langle \tau_{21} \rangle$, as shown in Fig.~\ref{fig:avgtau2comp} as a function of the groomed jet mass. All subjettiness variables are computed using the definitions of the axes in Ref.~\cite{Salam:2016yht}, and with $\beta = 1$.
The solid circles correspond to $\tau_{21}$ for groomed jets, and the dips are observed in much the same way as for $\langle N_2^1 \rangle$ in Fig.~\ref{fig:avgN2comp2}. The hollow circles correspond to $\tau_{21}$ computed from ungroomed variables, and the behaviour also follows a similar pattern as $\langle N_2^1 \rangle$ in Fig.~\ref{fig:avgN2comp2}. On the other hand, $\tau_{21}^{(\text{dic})}$, represented by crosses in the plot, receives a large enhancement at lower jet masses, so that part of the jet mass distribution would be strongly suppressed, even more than the high-mass peak.
In any case, we remark that the solution to the `bump running' effect should come from an appropriate tagging of these complex jets (using for example a generic anti-QCD tagger such as~\cite{Aguilar-Saavedra:2017rzt}) and a more robust jet grooming. Mixed tagging variables such as $\tau_{21}^{(\text{dic})}$ eliminate the effect discussed but at the expense of wiping out a possible multi-pronged jet signal across all the jet mass range.

\section{Dependence on grooming parameters}
\label{sec:b}

Grooming algorithms have parameters that control when soft contributions are removed or not. We investigate in this appendix the effect of changing these parameters in the soft drop algorithm for the $S \to A_1 A_2$ stealth boson signal. This algorithm reclusters the jet using the Cambridge-Aachen algorithm~\cite{Dokshitzer:1997in} to form a pairwise clustering tree. Afterwards, starting backwards the clustering procedure at the last subjet pair, the softer constituent is dropped unless the subjet pair is sufficiently `symmetric'. The condition for that is that the transverse momenta of these two subjets $p_{T1}$, $p_{T2}$ satisfy
\begin{equation}
\frac{\operatorname{min}(p_{T1},p_{T2})}{p_{T1}+p_{T2}} > z_\text{cut} \left( \frac{\Delta R_{12}}{R} \right)^\beta \,,
\end{equation}
with $R$ the jet radius and $\Delta R_{12}$ the lego-plot separation of the two subjets; $z_\text{cut}$ and $\beta$ are two free parameters that are adjusted to have a good grooming performance. If the condition is met, the groomed jet is defined by these two subjets; otherwise, the softer jet is dropped and the procedure is applied to the hardest one.

\begin{figure}[htb]
\begin{center}
\begin{tabular}{ccc}
\includegraphics[height=4.5cm,clip=]{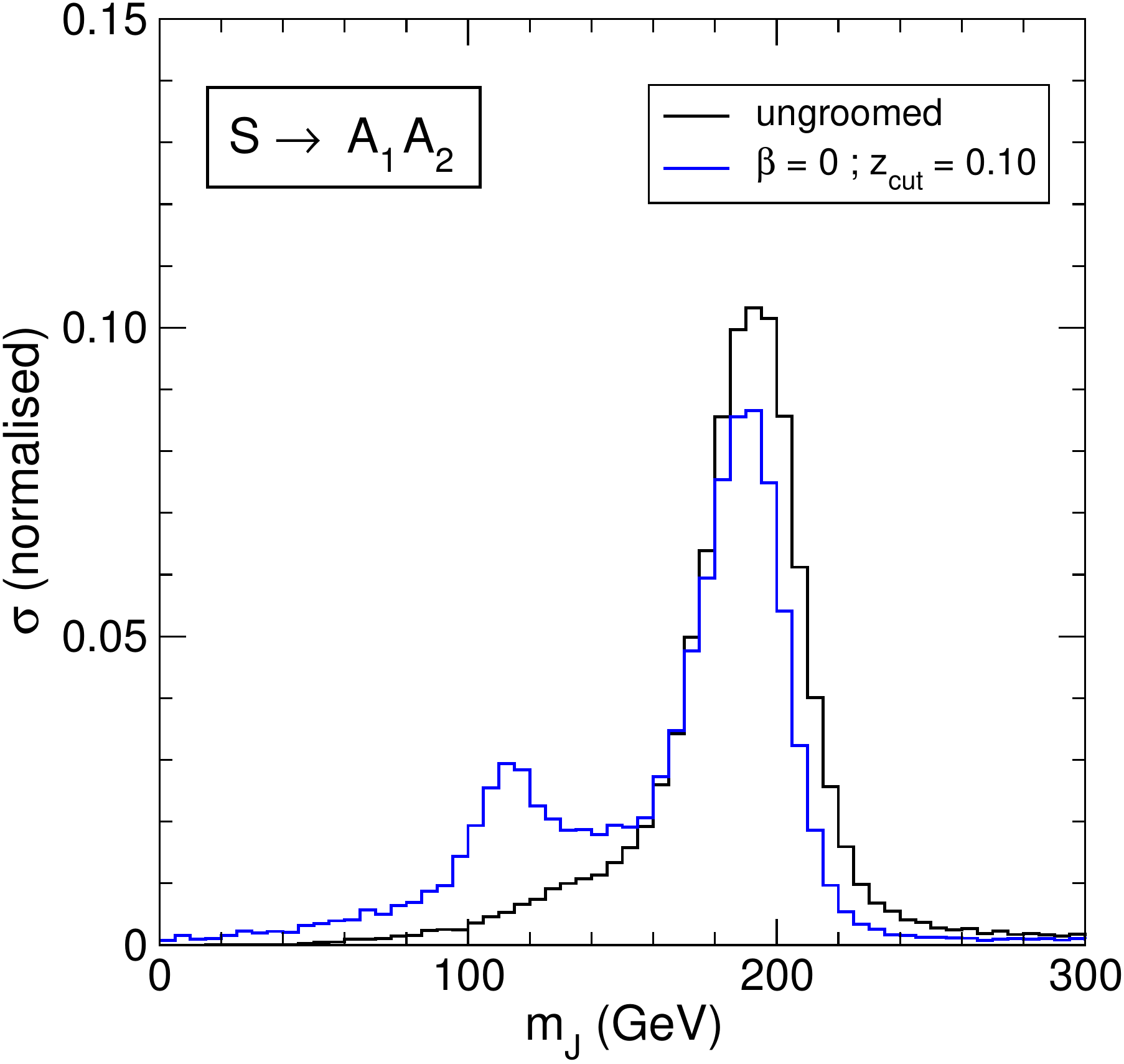} & 
\includegraphics[height=4.5cm,clip=]{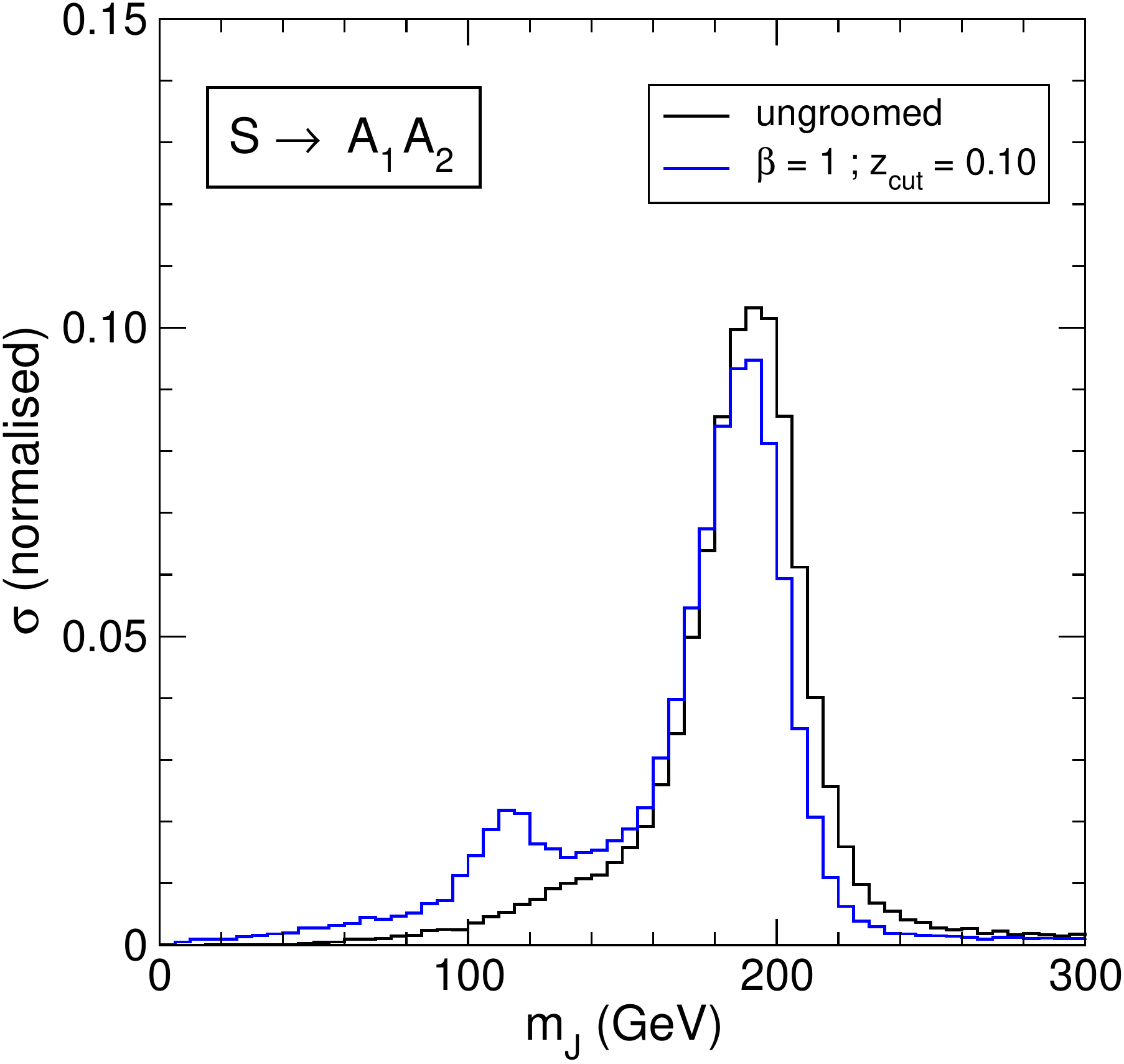} &
\includegraphics[height=4.5cm,clip=]{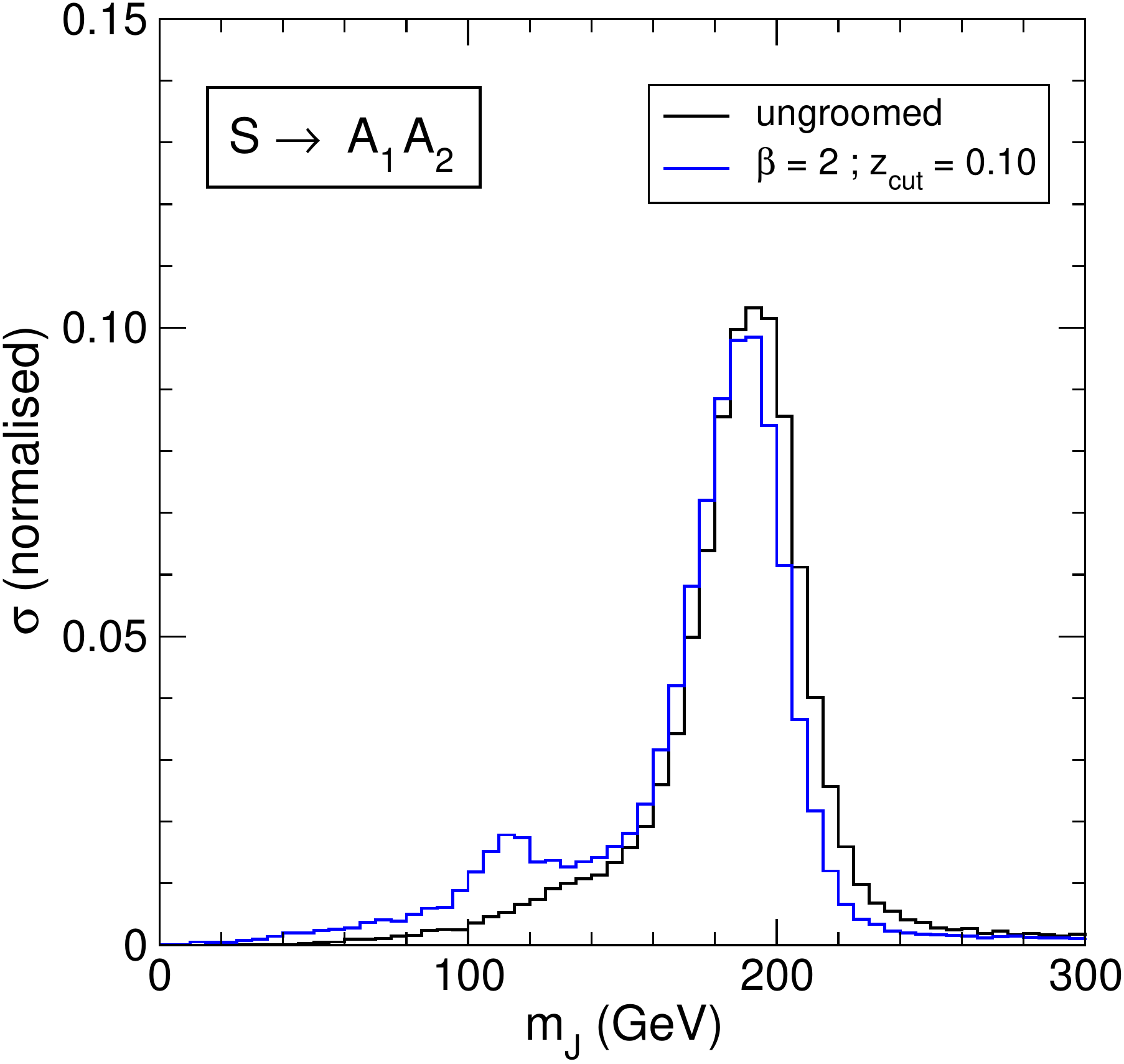} \\
\includegraphics[height=4.5cm,clip=]{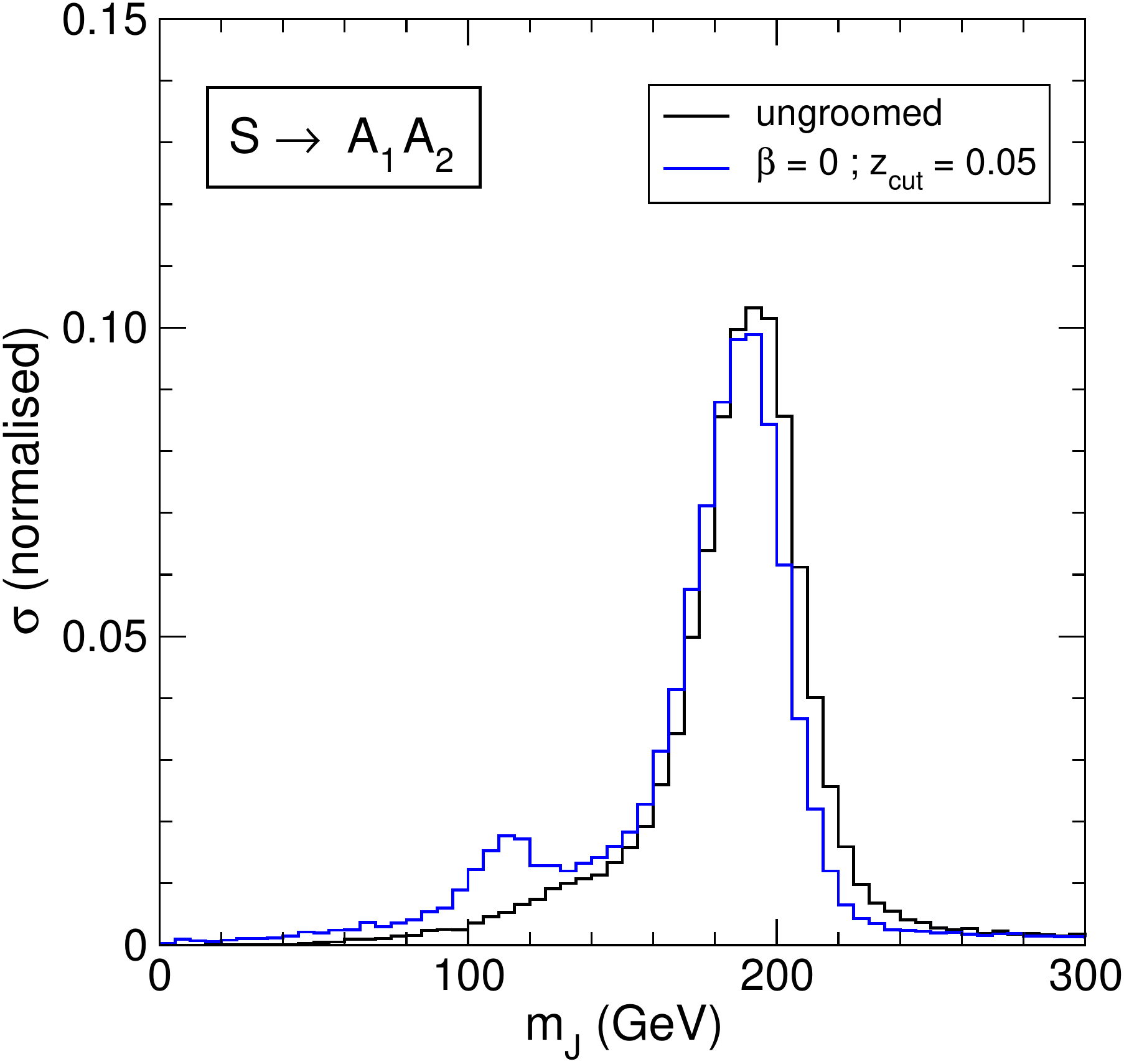} & 
\includegraphics[height=4.5cm,clip=]{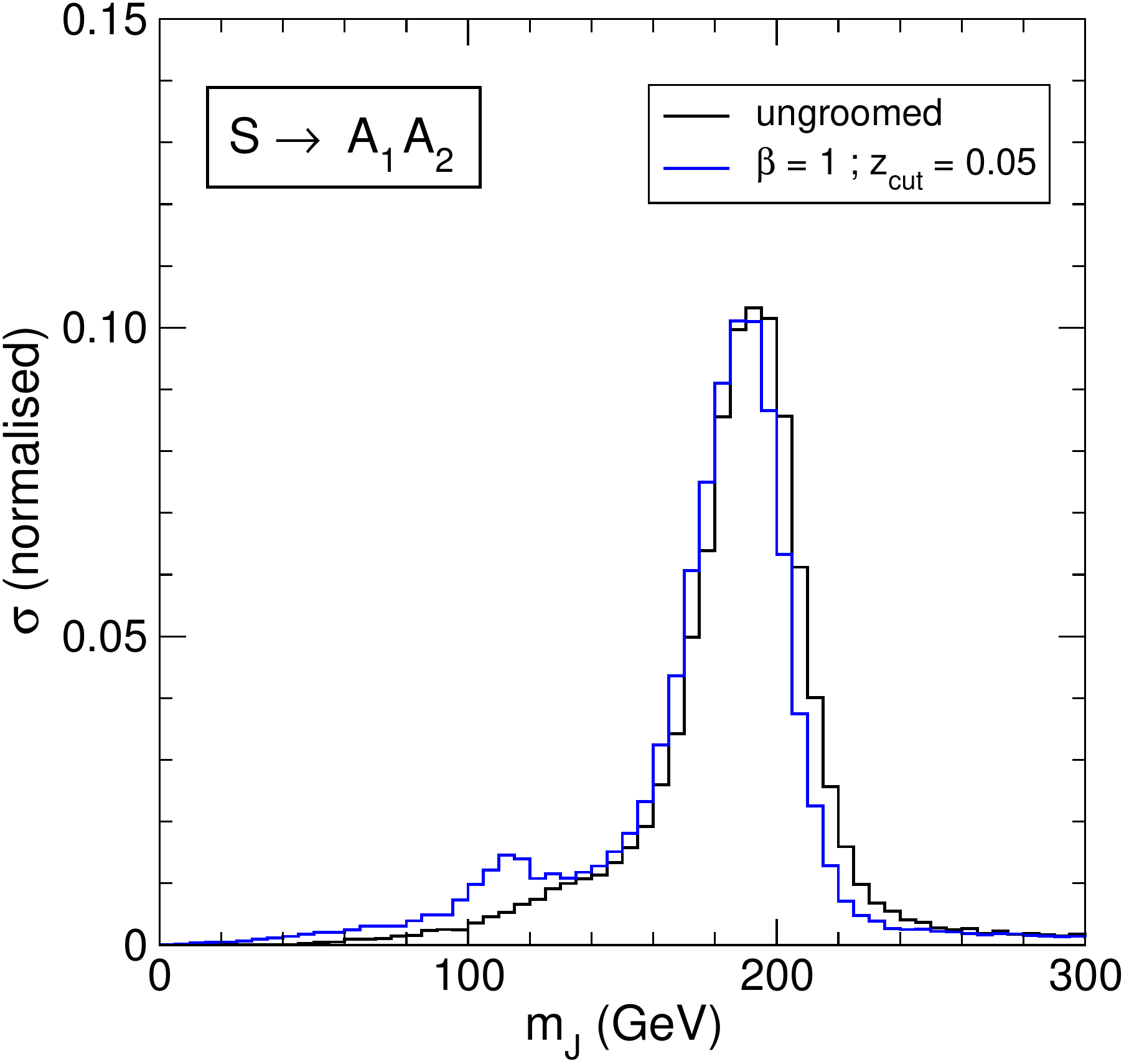} &
\includegraphics[height=4.5cm,clip=]{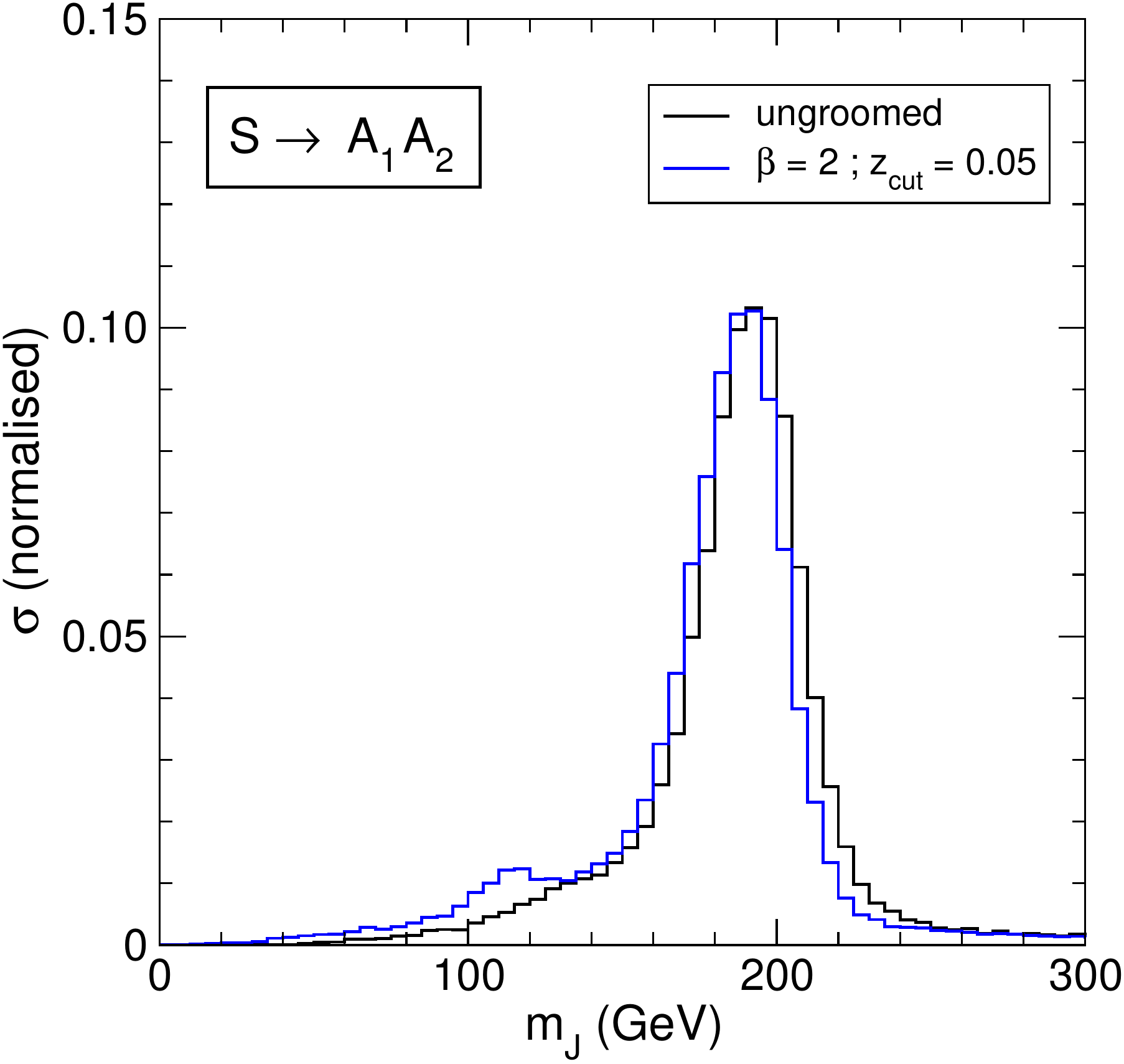} \\
\end{tabular}
\caption{Ungroomed jet mass (black) and groomed mass (blue) using the soft drop algorithm with different parameters, for the $S \to A_1 A_2$ stealth boson signal.}
\label{fig:tune1}
\end{center}
\end{figure}

\begin{figure}[htb]
\begin{center}
\begin{tabular}{ccc}
\includegraphics[height=4.5cm,clip=]{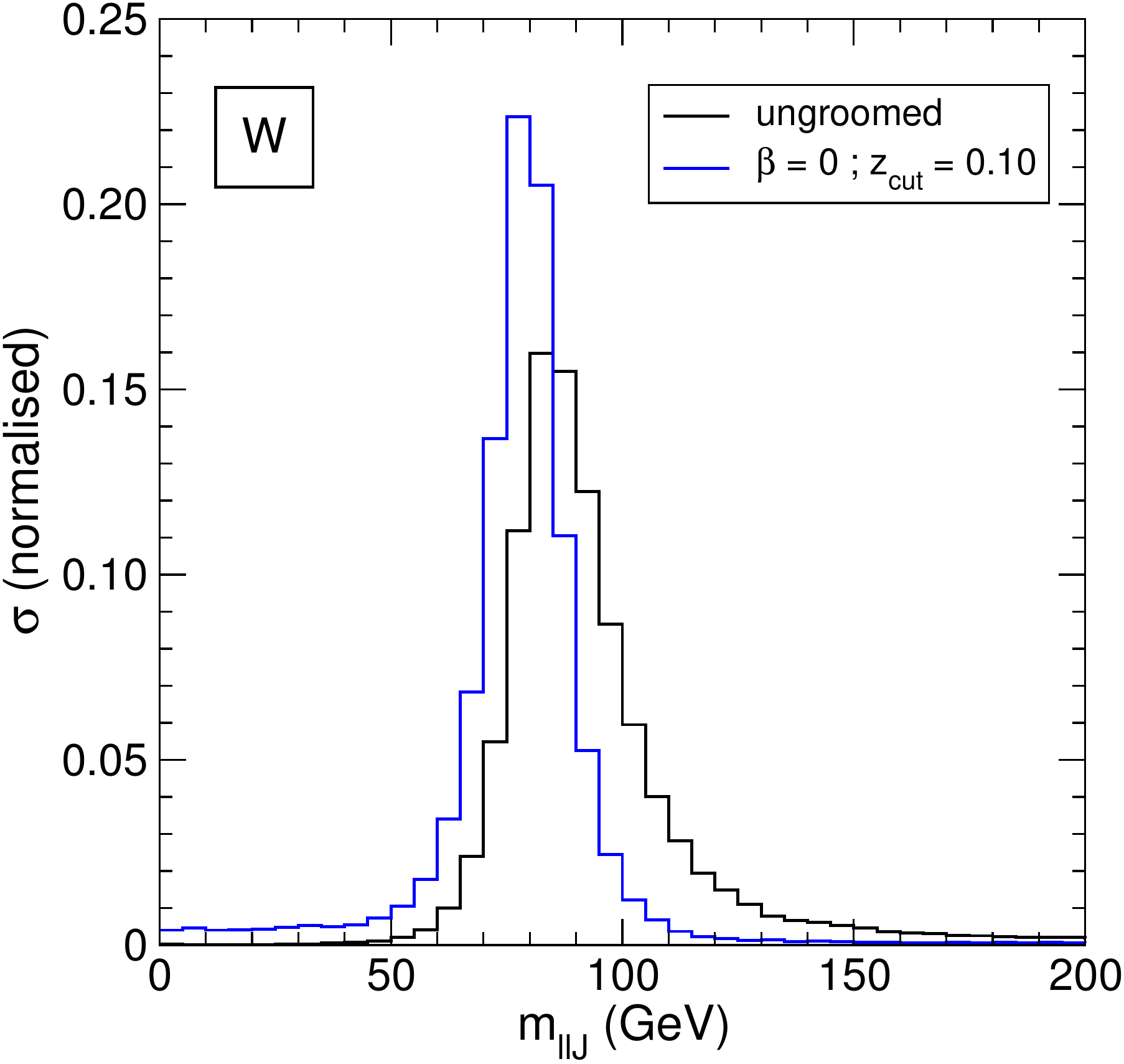} & 
\includegraphics[height=4.5cm,clip=]{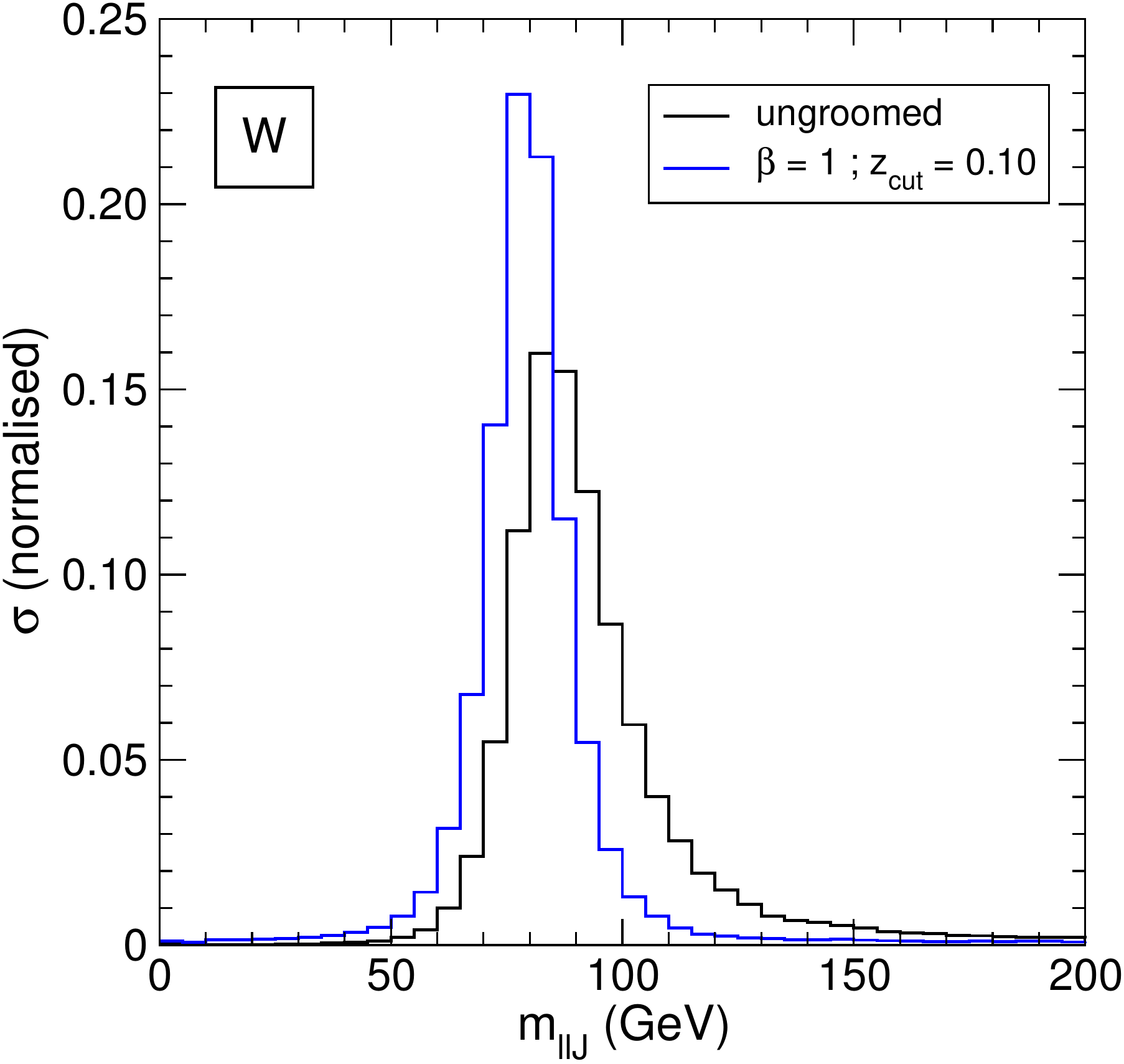} &
\includegraphics[height=4.5cm,clip=]{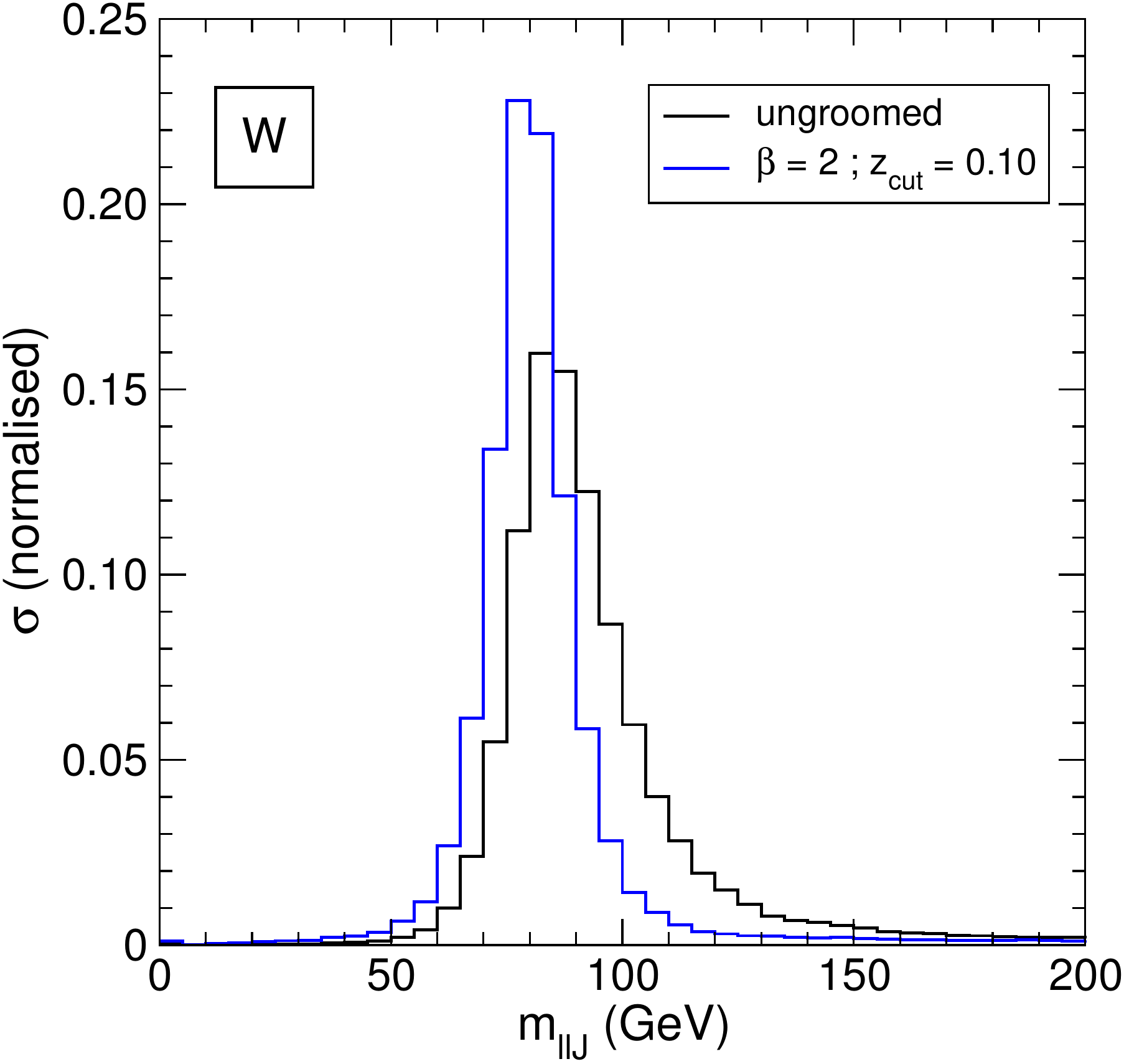} \\
\includegraphics[height=4.5cm,clip=]{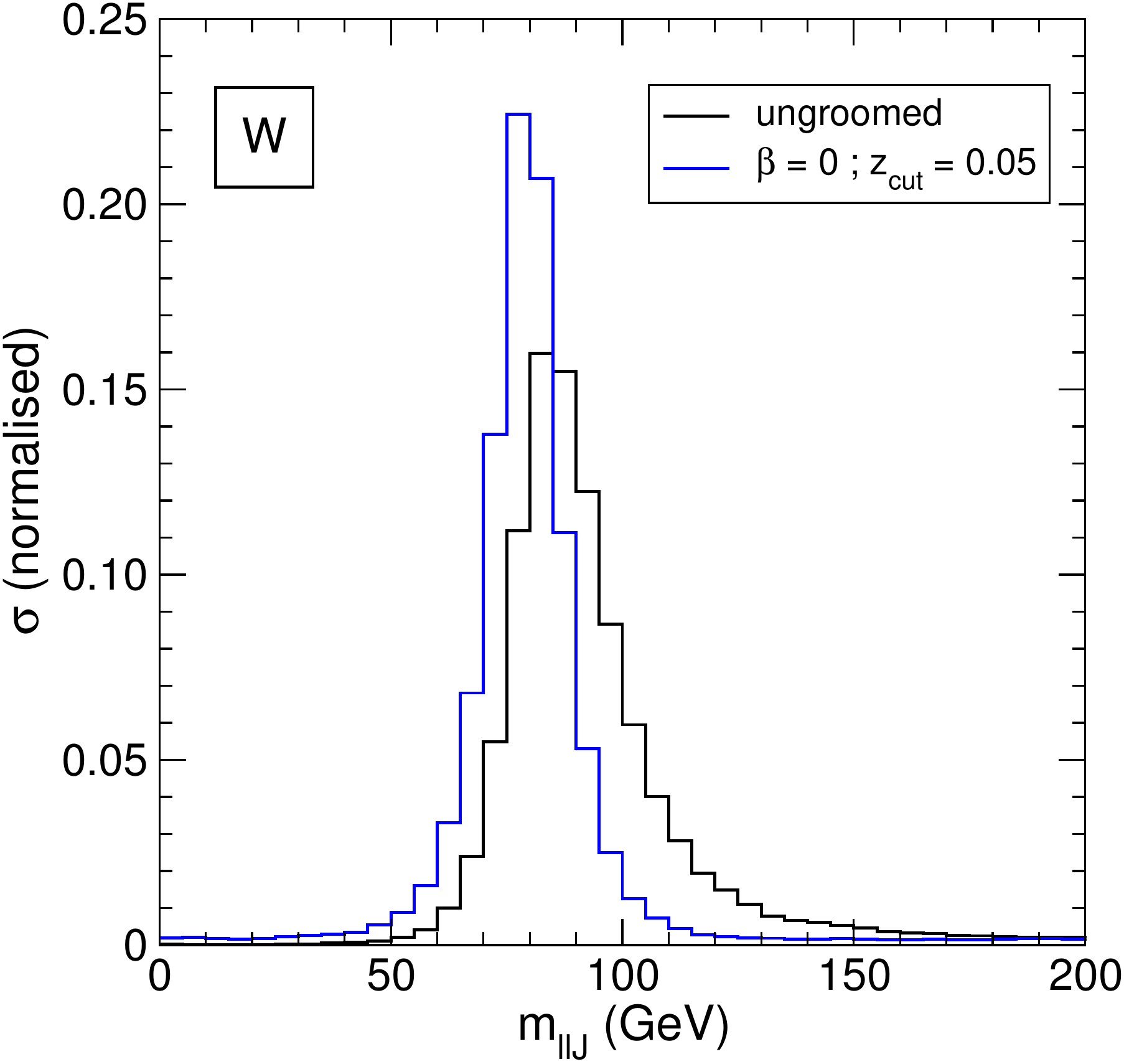} & 
\includegraphics[height=4.5cm,clip=]{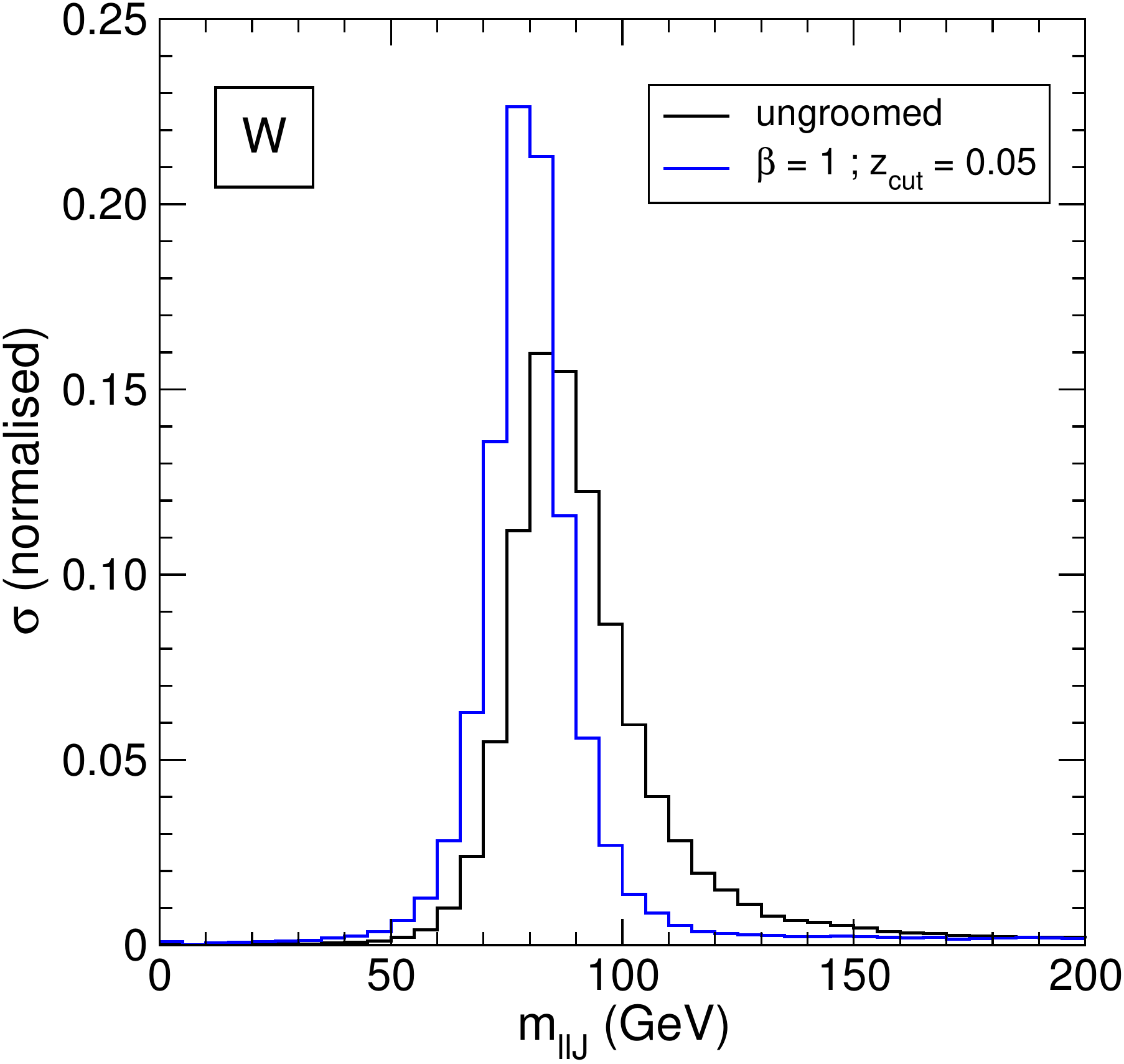} &
\includegraphics[height=4.5cm,clip=]{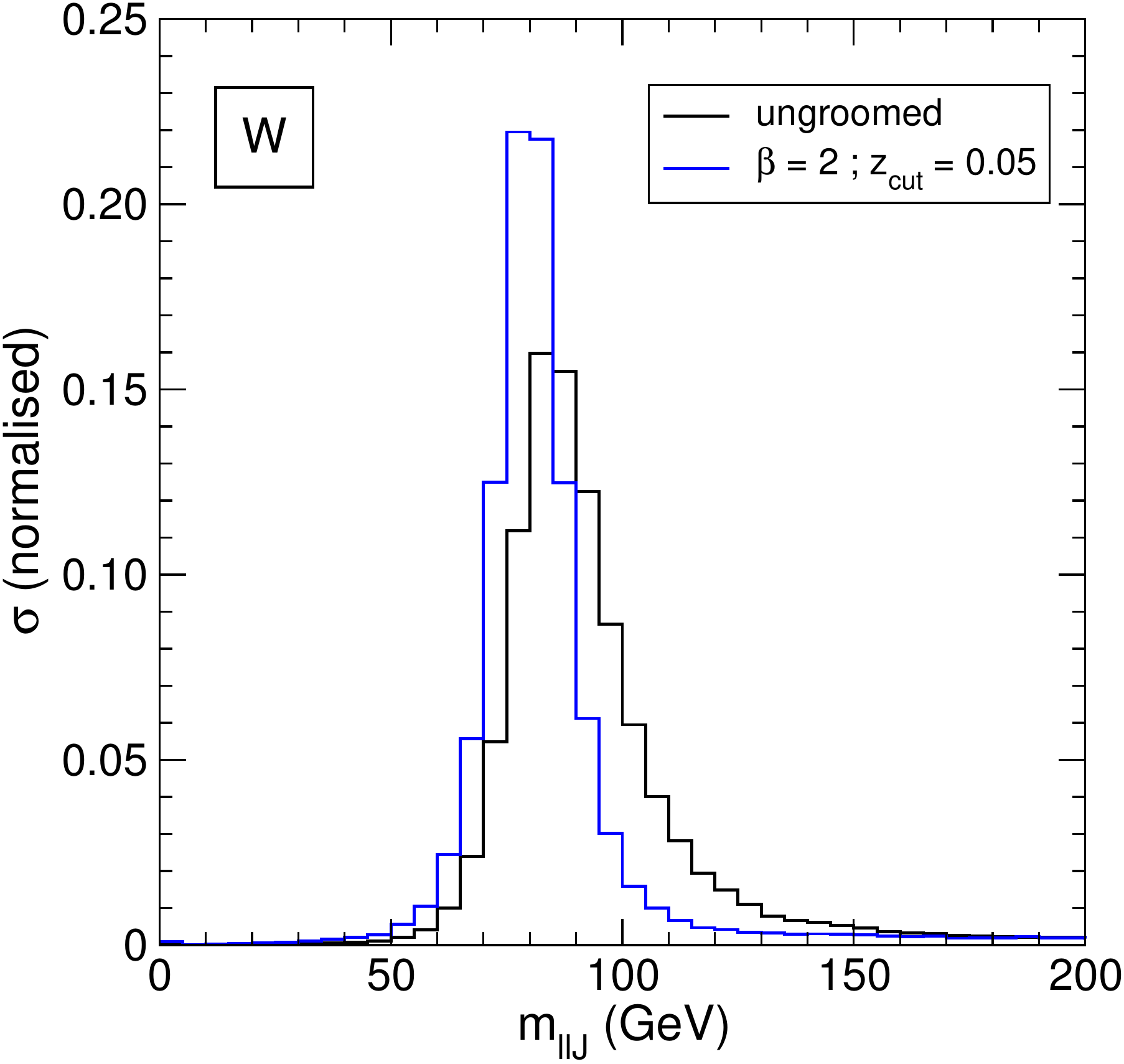} \\
\end{tabular}
\caption{Ungroomed jet mass (black) and groomed mass (blue) using the soft drop algorithm with different parameters, for a $W$ boson signal.}
\label{fig:tune2}
\end{center}
\end{figure}

In our analysis of sections~\ref{sec:2}--\ref{sec:5} we have used $z_\text{cut} = 0.1$, $\beta = 0$, which is a common choice and is actually adopted in the CMS analysis~\cite{Sirunyan:2017nvi}. We here explore some parameter combinations that make the groomer less aggresive, namely $z_\text{cut} = 0.05$ and $\beta = 1,2$. The results for the $S \to A_1 A_2$ stealth boson signal are presented in Fig.~\ref{fig:tune1}. For comparison, we show the results for a $W$ boson (i.e. the signal considered in section~\ref{sec:4.1}) in Fig.~\ref{fig:tune2}.

First of all, we remark that reducing the intensity of the grooming may constitute a problem in an environment with a high amount of pile-up such as the LHC Run 2. Therefore, an optimisation of the parameters for stealth boson signals sensitively depends on the pile-up present in each data taking period, and should be done with a more detailed simulation. With smaller $z_\text{cut}$ and/or larger $\beta$ the grooming is milder and, as expected, the size of the secondary bump decreases, as observed in Fig.~\ref{fig:tune1}. However, the resolution of the primary bump is slightly reduced too, which is an undesired effect, and the width of the peak is practically equal for the groomed and ungroomed jet. On the other hand, for $W$ bosons the grooming works well for the parameters explored, and only for $z_\text{cut} = 0.05$, $\beta = 2$ we can see in Fig.~\ref{fig:tune2} some degradation of the mass resolution.

\end{document}